\documentclass[prd, aps, showpacs]{revtex4}

\usepackage{latexsym, graphicx}

\begin{document}

\title{Temporal and Spatial Dependence of Quantum Entanglement\\
from a Field Theory Perspective}
\author{Shih-Yuin Lin}
\email{sylin@phys.cts.nthu.edu.tw}
\affiliation{Physics Division, National Center for Theoretical Sciences,
P.O. Box 2-131, Hsinchu 30013, Taiwan}
\author{B. L. Hu}
\email{blhu@umd.edu}
\affiliation{Joint Quantum Institute and Department of Physics,
University of Maryland, College Park, Maryland 20742-4111, USA}
\date{v1: December 23, 2008. v2: April 5, 2009}
\begin{abstract}
We consider the entanglement dynamics between two Unruh-DeWitt
detectors at rest separated at a distance $d$. This simple model when
analyzed properly in quantum field theory shows many interesting facets
and helps to dispel some misunderstandings of entanglement dynamics. 
We find that there is spatial dependence of quantum entanglement in
the stable regime due to the phase difference of vacuum fluctuations
the two detectors experience, together with the interference of the
mutual influences from the backreaction of one detector on the other.
When two initially entangled detectors are still outside each other's
light cone, the entanglement oscillates in time with an amplitude
dependent on spatial separation $d$. 
When the two detectors begin to have causal contact,
an interference pattern of the relative degree of entanglement
(compared to those at spatial infinity) develops a parametric
dependence on $d$. The detectors separated at those $d$ with a stronger
relative degree of entanglement enjoy longer disentanglement times.
In the cases with weak coupling and large separation, the detectors
always disentangle at late times. For sufficiently small $d$, the two
detectors can have residual entanglement even if they initially were
in a separable state, while for $d$ a little larger, there could be
transient entanglement created by mutual influences. However, we see 
no evidence of entanglement creation outside the light cone for 
initially separable states.
\end{abstract}


\maketitle

\section{Introduction}

Recently we have studied the disentanglement process between two
spatially separated Unruh-DeWitt (UD) detectors (pointlike objects
with internal degrees of freedom) or atoms, described by harmonic
oscillators, moving in a common quantum field: One at rest (Alice),
the other uniformly accelerating (Rob) \cite{LCH08}. These two
detectors are set to be entangled initially, while the initial state
of the field is the Minkowski vacuum. In all cases studied in
\cite{LCH08}, we obtain finite-time disentanglement (called ``sudden
death" of quantum entanglement \cite{YE04}), which are
coordinate dependent while the entanglement between the two detectors
at two spacetime points is independent of the choice of time slice
connecting these two events. Around the moment of complete
disentanglement there may be some short-time revival of entanglement
within a few periods of oscillations intrinsic to the detectors. In
the strong-coupling regime, the strong impact of vacuum fluctuations
experienced locally by each detector destroys their entanglement
right after the coupling is switched on.

In the above situation we find in \cite{LCH08} the event horizon for
the uniformly accelerated detector (Rob) cuts off the higher-order
corrections of mutual influences, and the asymmetric motions of Alice
and Rob obscure the dependence of the entanglement on the spatial
separation between them. To understand better how entanglement  
dynamics depends on the spatial separation between two quantum
objects, in this paper we consider the entanglement between two
detectors at rest separated at a distance $d$, possibly the simplest  
setup one could imagine. This will serve as a concrete model for us 
to investigate and explicate many subtle points and some essential
misconceptions related to quantum entanglement elicited by the
classic paper of Einstein-Podolsky-Rosen (EPR) \cite{EPR}.

\subsection{Entanglement at spacelike separation: quantum nonlocality?} 

One such misconception (or misnomer, for those who understand the
physics but connive to the use of the terminology) is ``quantum
nonlocality" used broadly and often too loosely in certain
communities \footnote{The issue of locality in quantum mechanics is
discussed in \cite{Unruh}. Note that in quantum information science
``quantum nonlocality" still respects causality \cite{PR97}.
In their classic paper \cite{EPR}, the EPR gedanken experiment was
introduced to bring out the incompleteness of quantum mechanics. EPR
made no mention of ``quantum nonlocality." This notion
seems to have crept in later for 
the situation when local measurements are performed at a spacelike
separated entangled pair, which cannot be described by any local
hidden variable theory \cite{Bell}.}. Some authors think that quantum
entanglement entails some kind of ``spooky action at a distance"
between two spacelike separated quantum entities (qubits, for
example), and may even extrapolate this to mean ``quantum
nonlocality." The phrase ``spooky action at a distance" when traced
to the source \cite{BE47} refers to the dependence of ``what really
exists at one event" on what kind of measurement is carried out at
the other, namely, the consequence of measuring one part of an
entangled pair. Without bringing in quantum measurement, one cannot
explore fully the existence or consequences of ``spooky action at a
distance" but one could still talk about quantum entanglement between
two spacelike separated qubits or detectors. This is the main theme
of our present investigation. We show in a simple and generic model
with calculations based on quantum field theory (QFT) that nontrivial
dynamics of entanglement outside the light cone does exist.

Another misconception is that entanglement set up between two localized 
quantum entities is independent of their spatial separation. This is false 
for open systems interacting with an environment \footnote{The environment 
here could be as innocuous and ubiquitous as a mediating quantum field or
vacuum fluctuations, whose intercession could in most cases engender
dissipative dynamics but in other special situations leave the
dynamics of the system unitary. For a discussion on the statistical
mechanical features of the equations of motion derived from a loop
expansion in quantum field theory, in particular the differences in
perspectives and results obtained from the in-in formulation in 
contradistinction to the in-out formulation, see, e.g., \cite{CH08}}. 
This has already been shown in two earlier investigations of the
authors \cite{ASH06, LCH08} and will be again in this paper.

A remark on nonlocality, or lack thereof, in QFT is in place here.
QFT is often regarded as ``local" in the sense that interactions of
the fields take place at the same spacetime point \footnote{In this
sense ``nonlocality" does exist in, e.g., noncommutative quantum field
theory or in certain quantum theories of spacetime, but that is a
much more severe breach of known physics, which need be dealt with at
a more fundamental level.}, e.g., for a bosonic field $\phi(x)$, a
local theory has no coupling of $\phi(x)$ and $\phi(y)$ at different
spacetime points $x$ and $y$. It follows that the vacuum expectation
value of the commutator $\left<\right. [\phi(x), \phi(y)]
\left.\right>$ vanishes for all $y$ outside the light cone of $x$,
which is what causality entails. Nevertheless, the Hadamard function
$\left<\right.\{\phi(x),\phi(y)\}\left.\right>$ is nonvanishing in
general, no matter $x-y$ is spacelike or timelike. In physical terms
the Hadamard function can be related to quantum noise in a stochastic
treatment of QFT \cite{HPZ}. In this restricted sense one could say
that QFT has certain nonlocal features. Of course it is well known
that in QFT processes occurring at spacelike separated events such as
virtual particle exchange are allowed.

\subsection{Issues addressed here}

With a careful and thorough analysis of this problem we are able to
address the following issues:

1) {\it Spatial separation between two detectors.}--Ficek and Tanas
\cite{FicTan06} as well as Anastopoulos, Shresta, and Hu (ASH)
\cite{ASH06} studied the problem of two spatially separated qubits
interacting with a common electromagnetic field. The former authors
while invoking the Born and Markov approximations find the appearance
of dark periods and revivals. ASH treat the non-Markovian behavior
without these approximations and find a different behavior at short
distances. In particular, for weak coupling, they obtain analytic
expressions for the dynamics of entanglement at a range of spatial
separation between the two qubits, which cannot be obtained when the
Born-Markov approximation is imposed.
A model with two detectors at rest in a quantum field at
finite temperature in (1+1)-dimensional spacetime has been considered
by Shiokawa in \cite{Tom08}, where some dependence of the early-time
entanglement dynamics on spatial separation can also be observed.

In \cite{LCH08} we did not see any simple proportionality between the
{\it initial} separation of Alice and Rob's detectors and the degree
of entanglement: The larger the separation, the weaker the entanglement
at some moments, but stronger at others. We wonder if this unclear
pattern arises because the spatial separation of the two detectors in
\cite{LCH08} changes in time and also in coordinate. In our present
problem the spatial separation between the two detectors is well
defined and remains constant in Minkowski time, so the dependence of
entanglement on the spatial separation should be much clearer and
distinctly identifiable.

2) {\it Stronger mutual influences.}--Among the cases we considered
in \cite{LCH08}, the largest correction from the mutual influences is
still under $2\%$ of the total while we have only the first and the
second-order corrections from the mutual influences. There the difficulty
for making progress is due to the complicated multidimensional
integrations in computing the back-and-forth propagations of the
backreactions sourced from the two detectors moving in different ways.
Here, for the case with both detectors at rest, the integration is
simpler and in some regimes we can include stronger and more higher-order
corrections of the mutual influences on the evolution of quantum entanglement.

3) {\it Creation of entanglement and residual entanglement.}--In
addition to finite-time disentanglement and the revival of quantum
entanglement for two detectors initially entangled, which have been
observed in \cite{LCH08} for a particular initial state, we expect to
see other kinds of entanglement dynamics with various initial states
and how it varies with spatial separations. Amongst the most
interesting behavior we found the creation of entanglement from an
initially separated state \cite{LHMC08} and the persistence of
residual entanglement at late times for two close-by detectors
\cite{PR07}.

\subsection{Summary of our findings}

When the mutual influences are sufficiently strong (under strong
coupling or small separation), the fluctuations of the detectors with
low natural frequency will accumulate, then get unstable and blow up.
As the separation approaches a merge distance (quantified later),
only for detectors with high enough natural frequencies will the
fluctuations not diverge eventually but acting more and more like
those in the two harmonic oscillator (2HO) quantum Brownian motion
(QBM) models \cite{CYH07, PR07} (where the two HOs occupy the same
spatial location) with renormalized frequencies.

If the duration of interaction is so short that each detector is
still outside the light cone of the other detector, namely, before the
first mutual influence reaches one another, the entanglement oscillates
in time with an amplitude dependent on spatial separation: At some
moments the larger the separation the weaker the entanglement,
but at other moments, the stronger the entanglement.
While such a behavior is affected by correlations of vacuum fluctuations 
locally experienced by the two detectors without causal contact,
there is no evidence for entanglement generation outside the
light cone suggested by Franson in Ref. \cite{Franson}. 

For an initially entangled pair of detectors, when one gets inside the
light cone of the other, certain interference patterns develop: At
distances where the interference is constructive the disentanglement
times are longer than those at other distances. This behavior is more
distinct when the mutual influences are negligible. For the detectors
separable initially, entanglement can be generated by mutual
influences if they are put close enough to each other.

At late times, under proper conditions, the detectors will be
entangled if the separation is sufficiently small, and separable
if the separation is greater than a specific finite distance. The
late-time behavior of the detectors is governed by vacuum fluctuations
of the field and independent of the initial state of the detectors.

Since the vacuum can be seen as the simplest medium that the two
detectors immersed in, we expect that the intuitions acquired here
will be useful in understanding quantum entanglement in atomic and
condensed matter systems (upon replacing the field in vacuum by those
in the medium). To this extent our results indicate that the
dependence of quantum entanglement on spatial separation of qubits
could enter in quantum gate operations (see \cite{ASH06} for comments
on possible experimental tests of this effect in cavity ions),
circuit layout, as well as having an effect on cluster states
instrumental to measurement-based quantum computing.

\subsection{Outline of this paper}

This paper is organized as follows. In Sec. II we describe our model
and the setup. In Sec. III the evolution of the operators is
calculated, then the instability for detectors with low natural
frequency is described in Sec. IV. We derive the zeroth-order results
in Sec. V, and the late-time results in Sec. VI. Examples with
different spatial separations of detectors in the weak-coupling limit
are given in Sec. VII. We conclude with some discussions in Sec. VIII.
A late-time analysis on the mode functions is performed in Appendix A,
while an early-time analysis of the entanglement dynamics in the
weak-coupling limit is given in Appendix B.

\section{The model}

Let us consider the Unruh-DeWitt detector theory in (3+1)-dimensional
Minkowski space described by the action \cite{LCH08, LH2005}
\begin{eqnarray}
  S &=& -\int d^4 x {1\over 2}\partial_\mu\Phi\partial^\mu\Phi  + 
    \sum_{j=A,B}\left\{\int d\tau_j {1\over 2}\left[\left(\partial_{\tau_j}
     Q_j\right)^2 -\Omega_0^2 Q_j^2\right] + \lambda_0\int d^4 x \Phi (x) 
    \int d\tau_j Q_j(\tau_j)\delta^4\left(x^{\mu}-z_j^{\mu}(\tau_j)\right)
    \right\}, \label{Stot1}
\end{eqnarray}
where the scalar field $\Phi$ is assumed to be massless, and $\lambda_0$
is the coupling constant. $Q_A$ and $Q_B$ are the internal degrees of
freedom of the two detectors, assumed to be two identical harmonic
oscillators with mass $m_0 =1$, bare natural frequency $\Omega_0$, and
the same local time resolution so their cutoffs 
in two-point functions \cite{LH2005} are the same. The left detector is
at rest along the world line $z_A^\mu(t)=(t,-d/2,0,0)$ and the right
detector is sitting along $z_B^\mu(t)=( t,d/2,0,0)$. The proper times for
$Q_A$ and $Q_B$ are both the Minkowski time, namely, $\tau_A=\tau_B=t$.

We assume at $t=0$ the initial state of the combined system is a
direct product of the Minkowski vacuum $\left|\right. 0_M \left.
\right>$ for the field $\Phi$ and a quantum state $\left|\right. Q_A,
Q_B \left.\right>$ for the detectors $Q_A$ and $Q_B$, taken to be a
squeezed Gaussian state with minimal uncertainty, represented by the
Wigner function of the form
\begin{equation}
  \rho(Q_A,P_A,Q_B,P_B) = {1\over \pi^2\hbar^2}\exp -{1\over 2}\left[
  {\beta^2\over\hbar^2}\left( Q_A + Q_B\right)^2 +
  {1\over \alpha^2}\left( Q_A - Q_B\right)^2 +
  {\alpha^2\over\hbar^2}\left( P_A - P_B\right)^2 +
  {1\over \beta^2}\left( P_A + P_B\right)^2 \right].
\label{initGauss}
\end{equation}
How the two detectors are initially entangled is determined by
properly choosing the parameters $\alpha$ and $\beta$ in $Q_A$ and
$Q_B$. When $\beta^2 = \hbar^2/\alpha^2$, the Wigner function
$(\ref{initGauss})$ becomes a product of the Wigner functions for 
$Q_A$, $P_A$ and for $Q_B$, $P_B$, thus separable. If one
further chooses $\alpha^2 = \hbar/\Omega$, then the Wigner function 
will be initially in the ground state of the two free detectors. 

After $t=0$ the coupling with the field is turned on and the
detectors begin to interact with each other through the field while
the reduced density matrix for the two detectors becomes a mixed state.
The linearity of $(\ref{Stot1})$ guarantees that the quantum state
of the detectors is always Gaussian. Thus the dynamics of quantum
entanglement can be studied by examining the behavior of the quantity
$\Sigma$ \cite{LCH08} and the logarithmic negativity $E_{\cal N}$
\cite{VW02}:
\begin{eqnarray}
  \Sigma &\equiv&\det\left[ {\bf V}^{PT}+{i\hbar\over 2}{\bf M}\right],\\
  E_{\cal N} &\equiv& \max \left\{ 0, -\log_2 2c_- \right\}.
\end{eqnarray}
Here ${\bf M}$ is the symplectic matrix ${\bf 1}\otimes (-i)\sigma_y$,
${\bf V}^{PT}$ is the partial transpose ($(Q_A, P_A, Q_B, P_B)\to
(Q_A, P_A, Q_B, -P_B)$) of the covariance matrix
\begin{equation}
  {\bf V} = \left( \begin{array}{cc} {\bf v}_{AA} & {\bf v}_{AB} \\
                   {\bf v}_{BA} & {\bf v}_{BB} \end{array}\right)
\end{equation}
in which the elements of the $2\times 2$ matrices ${\bf v}_{ij}$ are
symmetrized two-point correlators ${\bf v}_{ij}{}^{mn} =
\left<\right.{\cal R}_i^m , {\cal R}_j^n \left.\right> \equiv
\left<\right.({\cal R}_i^m {\cal R}_j^n + {\cal R}_j^n {\cal R}_i^m )
\left.\right>/2$ with ${\cal R}_i^m = (Q_i(t), P_i(t))$, $m,n= 1,2$
and $i, j = A,B$. $(c_+, c_-)$ is the symplectic spectrum of
${\bf V}^{PT}+ (i\hbar/2){\bf M}$, given by
\begin{equation}
  c_\pm \equiv \left[Z \pm \sqrt{Z^2-4\det {\bf V}}\over 2
    \right]^{1/2}
\label{SympSpec}
\end{equation}
with
\begin{equation}
  Z = \det {\bf v}_{AA} + \det {\bf v}_{BB} - 2 \det {\bf v}_{AB}.
\end{equation}

For the detectors in Gaussian state, $E_{\cal N}>0$, $\Sigma <0$,
and $c_- < \hbar/2$, if and only if the quantum state of the detectors
is entangled \cite{Si00}.
$E_{\cal N}$ is an entanglement monotone \cite{Plenio05} whose value
can indicate the degree of entanglement: below we say the two detectors
have a stronger entanglement if the associated $E_{\cal N}$ is greater.
In the cases considered in Ref. \cite{LCH08} and this paper, the
behavior of $\Sigma$ is similar to $-E_{\cal N}$ when it is nonzero.
Indeed, the quantity $\Sigma$ can also be written as
\begin{equation}
  \Sigma = \left( c_+^2 -{\hbar^2\over 4}\right)\left( c_-^2 -
    {\hbar^2\over 4}\right)= \det {\bf V} - {\hbar^2 \over 4}Z +
    {\hbar^4\over 16}.
\end{equation}
We found it is more convenient to use $\Sigma$ in calculating the
disentanglement time. We also define the uncertainty function
\begin{equation}
  \Upsilon  \equiv \det \left[ {\bf V} + i{\hbar\over 2}{\bf M}
  \right],   \label{Uncert}
\end{equation}
so that $\Upsilon \ge 0$ is the uncertainty relation \cite{Si00}.

To obtain these quantities, we have to know the correlators
$\left<\right.{\cal R}_i^m ,{\cal R}_j^n \left.\right>$, so we are 
calculating the evolution of operators ${\cal R}_i^m$ in the following.

\section{Evolution of operators}
\label{EvoOp}

Since the combined system $(\ref{Stot1})$ is linear, in the Heisenberg
picture \cite{LH2005, LH2006}, the operators evolve as 
\begin{eqnarray}
  \hat{Q}_i(t) &=& \sqrt{\hbar\over 2\Omega_r}\sum_j\left[
    q_i^{(j)}(t)\hat{a}_j +q_i^{(j)*}(t)\hat{a}_j^\dagger \right]
   +\int {d^3 k\over (2\pi)^3}\sqrt{\hbar\over 2\omega}
    \left[q_i^{(+)}(t,{\bf k})\hat{b}_{\bf k} +
    q_i^{(-)}(t,{\bf k})\hat{b}_{\bf k}^\dagger\right], \\
  \hat{\Phi}(x) &=& \sqrt{\hbar\over 2\Omega_r}\sum_j\left[
    f^{(j)}(x)\hat{a}_j^{}+f^{(j)*}(x)\hat{a}_j^\dagger \right]
    +\int {d^3 k\over (2\pi)^3}
    \sqrt{\hbar\over 2\omega}\left[f^{(+)}(x,{\bf k})\hat{b}_{\bf k}
    +f^{(-)}(x,{\bf k})\hat{b}_{\bf k}^\dagger\right],
\end{eqnarray}
with $i,j = A,B$. $q_i^{(j)}$, $q_i^{(\pm)}$, $f^{(j)}$, and
$f^{(\pm)}$ are the (c-number) mode functions, $\hat{a}_j$ and
$\hat{a}_j^\dagger$ are the lowering and raising operators for the 
free detector $j$, while $\hat{b}_{\bf k}$ and $\hat{b}_{\bf k}^\dagger$
are the annihilation and creation operators for the free field. 
The conjugate momenta are $\hat{P}_j(t) =\partial_t\hat{Q}_j(t)$ and
$\hat{\Pi}(x)=\partial_t\hat{\Phi}(x)$. The evolution equations for the
mode functions have been given in Eqs.$(9)-(12)$ in Ref. \cite{LCH08}
with $z_A(t)$ and $z_B(\tau)$ there replaced by $z_A^\mu(t)=(t,-d/2,0,0)$ 
and $z_B^\mu(t)=(t,d/2,0,0)$ here. Since we have assumed that the two 
detectors have the same frequency cutoffs in their local frames, one
can do the same renormalization on frequency and obtain their effective
equations of motion under the influence of the quantum field \cite{LH2005}:
\begin{eqnarray}
  \left( \partial_t^2 +2\gamma\partial_t + \Omega_r^2 \right)q_i^{(j)}(t)
    &=& {2\gamma\over d}\theta (t-d) \bar{q}_i^{(j)}( t-d),\label{eomqA2} \\
  \left( \partial_t^2 +2\gamma\partial_t + \Omega_r^2 \right)
    q_i^{(+)}(t,{\bf k}) &=& {2\gamma\over d}\theta(t-d) \bar{q}_i^{(+)}
    (t-d,{\bf k})+\lambda_0 f_0^{(+)}(z_i(t),{\bf k}), \label{eomqA+2}
\end{eqnarray}
where $\bar{q}_B \equiv q_A$, $\bar{q}_A \equiv q_B$, $\Omega_r$ is
the renormalized frequency obtained by absorbing the singular behavior 
of the retarded solutions for $f^{(j)}$ and $f^{(\pm)}$ around 
their sources (for details, see Sec.IIA in Ref.\cite{LH2005}).
Also $\gamma \equiv \lambda_0^2/8\pi$, and
$f_0^{(+)}(x,{\bf k}) \equiv e^{-i \omega t + i{\bf k\cdot x}}$,
with $\omega=|{\bf k}|$. Here one can see that $q_B$ and $q_A$ are
affecting, and being affected by, each other causally with a
retardation time $d$.

The solutions for $q_i^{(j)}$ and $q_i^{(+)}$ satisfying the initial
conditions
$f^{(+)}(0,{\bf x};{\bf k}) = e^{i{\bf k\cdot x}}$,
$\partial_t f^{(+)}(0,{\bf x};{\bf k})=-i\omega e^{i{\bf k\cdot x}}$,
$q_j^{(j)}(0) =1$, $\partial_t q_j^{(j)}(0)= -i\Omega_r$,
and $f_i^{(j)} (0,{\bf x}) =\partial_t f_i^{(j)} (0,{\bf x}) =
q^{(+)}(0;{\bf k})= \partial_t q^{(+)}(0;{\bf k}) = \bar{q}_j^{(j)}(0) =
\partial_t\bar{q}_j^{(j)}(0)=0$ (no summation over $j$) are
\begin{eqnarray}
  q_{j}^{(+)}({\rm k};t) &=& {\sqrt{8\pi\gamma}\over \Omega}
    \sum_{n=0}^\infty \theta(t-nd)\left( 2\gamma\over\Omega d\right)^n
    e^{(-1)^n i k_1 z^1_j} \left\{ (M_1 - M_2)^{n+1}
    e^{-i\omega(t-nd)} + \right.\nonumber\\& & \left.
    e^{-\gamma(t-nd)}\sum_{m=0}^n (M_1-M_2)^{n-m} \left[M_2 W_m(t-nd) -
    M_1 W_m^*(t-nd)\right]\right\},
\label{qjp}
\end{eqnarray} and
\begin{equation}
  q_j^{(j)}=\sum_{n=0}^\infty q_{2n}, \,\,\,\,\,
  \bar{q}_j^{(j)}=\sum_{n=0}^\infty q_{2n+1}  \label{qjj}
\end{equation}
(no summation over $j$), where $\Omega \equiv \sqrt{\Omega_r^2 - \gamma^2}$,
$M_1 \equiv (-\omega-i\gamma +\Omega)^{-1}$,
$M_2 \equiv (-\omega-i\gamma -\Omega)^{-1}$,
$W_0(t) \equiv e^{i\Omega t}$,
\begin{equation}
  W_n(t) \equiv \int_0^t dt_{n-1} \sin\Omega(t-t_{n-1}) \int_0^{t_{n-1}}
    dt_{n-2}\sin\Omega(t_{n-1}-t_{n-2}) \cdots \int_0^{t_1} dt_0
    \sin\Omega(t_1-t_0) W_0(t_0),
\end{equation}
for $n\ge 1$, and
\begin{equation}
  q_n(t) = \theta(t-nd)\left( 2\gamma\over\Omega d\right)^n
    e^{-\gamma(t-nd)} \left[s_1 W_n(t-nd)+s_2 W_n^*(t-nd)\right],
\end{equation}
with $s_1 \equiv [1 - \Omega^{-1}(\Omega_r+i \gamma)]/2$,
and $s_2 \equiv [1 + \Omega^{-1}(\Omega_r+i \gamma)]/2$.

Using the mode functions Eqs. $(\ref{qjp})$ and $(\ref{qjj})$
one can calculate the correlators of the detectors for the covariance
matrix ${\bf V}$ \cite{LCH08}, each splitting into two parts ($\left<\right.
..\left.\right>_{\rm a}$ and $\left<\right. .. \left.\right>_{\rm v}$)
due to the factorized initial state. Because of symmetry, one has
$\left<\right.Q_A^2 \left.\right>= \left<\right.Q_B^2 \left.\right>$,
$\left<\right.P_A^2 \left.\right>= \left<\right.P_B^2 \left.\right>$,
and $\left<\right.Q_A, P_B \left.\right>= \left<\right.Q_B, P_A
\left.\right>$. So only six two-point functions need to be calculated
for ${\bf V}$.

Since $q_n \sim [\gamma (t-n d)/\Omega d]^n e^{-\gamma(t-n d)}/n!$ for
large $t$, $q_n$ will reach its maximum amplitude ($\approx (n/e \Omega
d)^n/n!$) around $t-n d\approx n/\gamma$, which makes the numerical
error of the long-time behavior of ${\bf V}$ difficult to control.
Fortunately for the late-time behavior for all $d$ and the long-time
behavior for very small or very large $d$, we still have good
approximations, as we shall see below. However, before we proceed, 
the issue of instability should be addressed first.

\section{Instability of low-frequency harmonic oscillators}
\label{instab}

Combining the equations of motion for $q_A^{(A)}$ and $q_B^{(A)}$, one has
\begin{equation}
  \left(\partial_t^2 +2\gamma\partial_t +\Omega_r^2 \right)q_{\pm}^{(A)}(t)
    = \pm {2\gamma\over d} q_{\pm}^{(A)}(t-d). \label{eomqpm}
\end{equation}
where $q_{\pm}^{(A)}(t) \equiv q_A^{(A)}(t)\pm q_B^{(A)}(t)$. For $t>d$ and
when $d$ is small, one may expand $q_{\pm}^{(A)}(t-d)$ around $t$ so that
\begin{equation}
  \left(\partial_t^2 +2\gamma\partial_t +\Omega_r^2\right) q_{\pm}^{(A)}(t)
    = \pm {2\gamma\over d}\left[ q_{\pm}^{(A)}(t)
     - d \partial_t q_{\pm}^{(A)}(t)
     + {d^2\over 2} \partial_t^2 q_{\pm}^{(A)}(t)
     - {d^3\over 3!} \partial_t^3 q_{\pm}^{(A)}(t) +\cdots\right],
\label{eomsmalld}
\end{equation}
or
\begin{eqnarray}
  \left[\partial_t^2 +4\gamma\partial_t +\left(\Omega_r^2 -
  {2\gamma\over d} \right)\right] q_+^{(A)}(t)= O(\gamma d),\label{EOMqR+}\\
  \left[\partial_t^2 +\left(\Omega_r^2+{2\gamma\over d} \right)\right]
    q_-^{(A)}(t) = O(\gamma d).  \label{EOMqR-}
\end{eqnarray}
If we start with a small renormalized frequency $\Omega_r$ and a
small spatial separation $d <  2\gamma/\Omega_r^2$ with $\gamma d$
kept small so the $O(\gamma d)$ terms can be neglected, then $q_+^{(A)}$
will be exponentially growing since its effective frequency becomes
imaginary ($\Omega_r^2 - (2\gamma/d) < 0$), while $q_-^{(A)}$ oscillates
without damping. A similar argument shows that $q_{\pm}^{(B)}$
will have the same instability when two harmonic oscillators
with small $\Omega_r^2$ are situated close enough to each other.

One may wonder whether the $O(\gamma d)$ terms can alter the above
observations. In Appendix \ref{LateAna} we perform a late-time
analysis, which shows the same instability.
The conclusion is, if $\Omega_r^2 < 2\gamma/d$, all the mode functions will
grow exponentially in time so the correlators $\left<\right.Q_i,Q_j\left.
\right>$ or the quantum fluctuations of the detectors diverge at late times.
Accordingly, we define
\begin{equation}
    d_{ins}\equiv 2\gamma/\Omega_r^2 \label{dins}
\end{equation}
as the ``radius of instability." For two detectors with separation
$d> d_{ins}$, the system is stable. For the cases with $d = d_{ins}$,
a constant solution for $q_+^{(j)}$ at late times is acquired by
$(\ref{EOMqR+})$, while for $d < d_{ins}$, the system is unstable.

Below we restrict our discussion to the stable regime,
$\Omega_r^2 >2\gamma/d$.

\section{Zeroth-order results}
\label{ZOR}

\begin{figure}
\includegraphics[width=9cm]{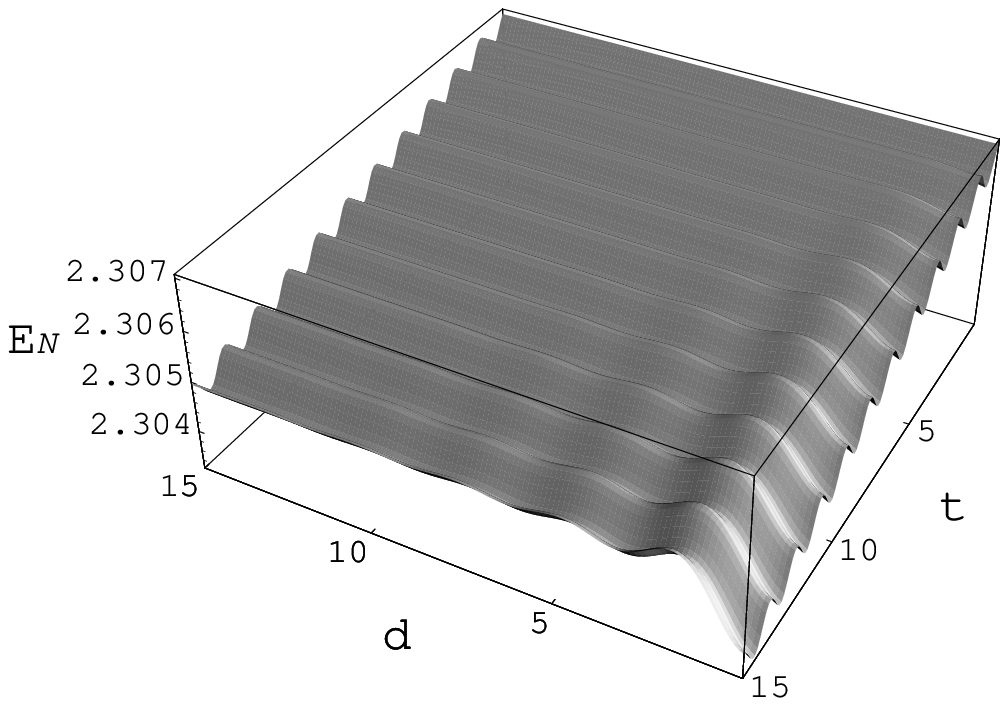}
\includegraphics[width=7cm]{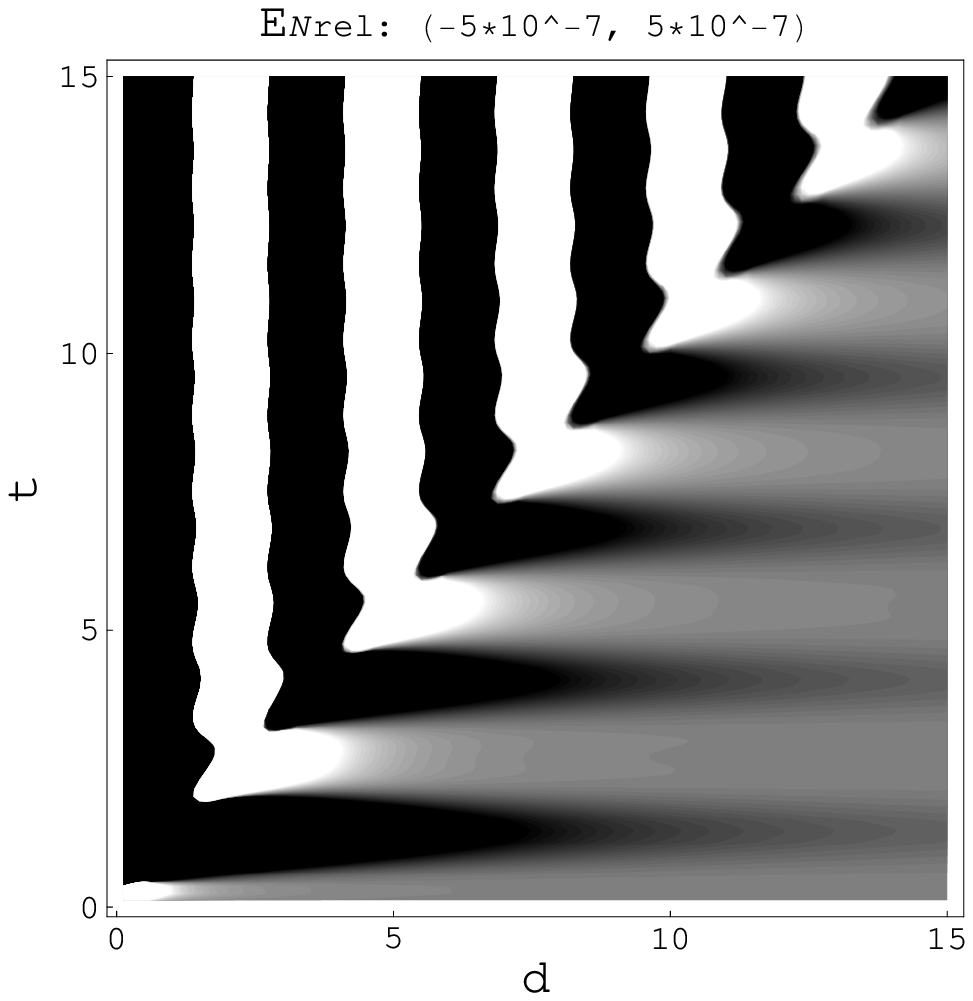}\\  
\includegraphics[width=8cm]{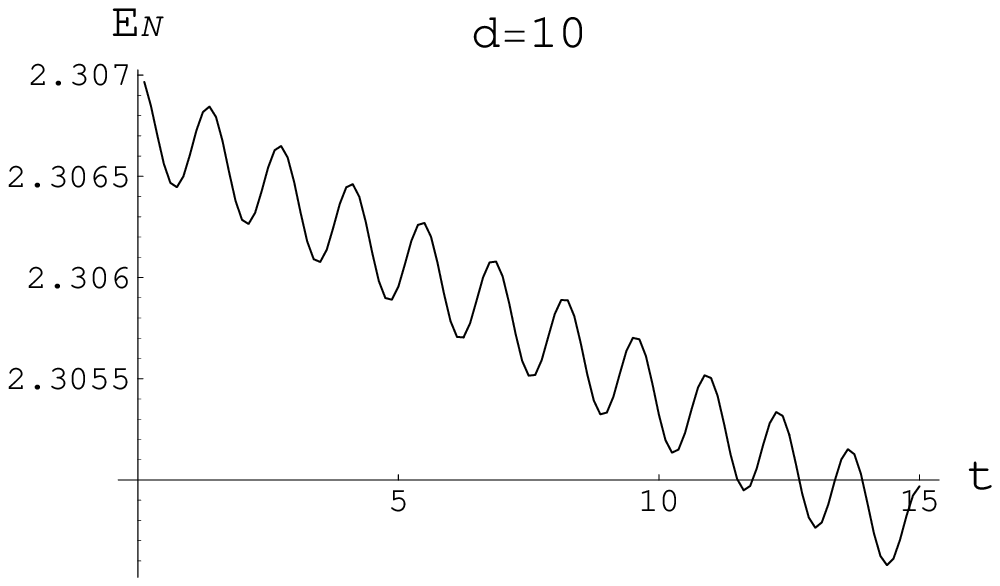}
\includegraphics[width=8cm]{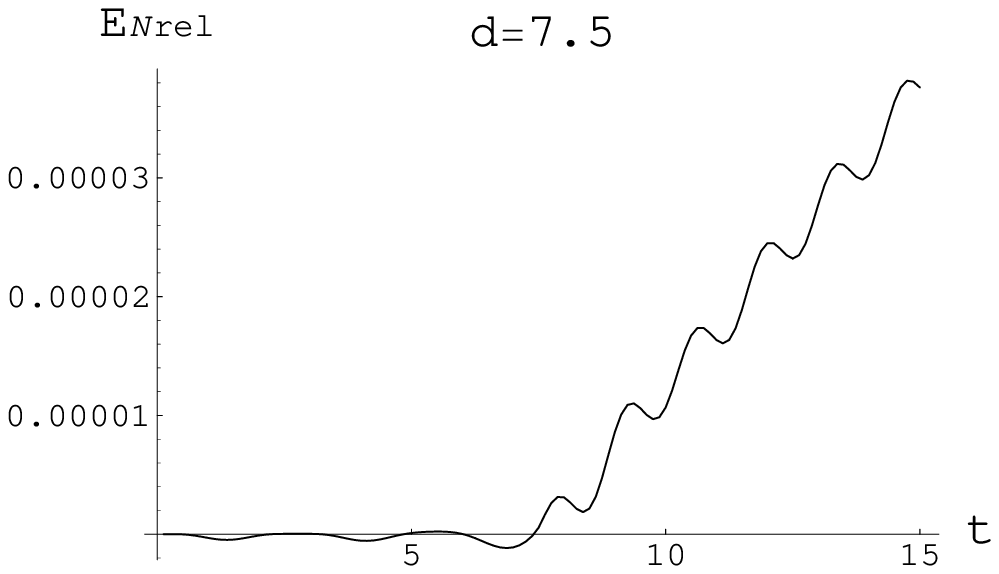}
\caption{The zeroth-order results, no mutual influence is included here.
$\gamma=10^{-5}$, $\Omega=2.3$, $\Lambda_0=\Lambda_1=20$, and
$(\alpha,\beta)=(1.1, 4.5)$.
The two plots on the left are for the zeroth-order $E_{\cal N}^{(0)}$
which is seen to decrease and disentangle in time. (The behavior of
$\Sigma^{(0)}$ is similar to $-E_{\cal N}^{(0)}$ but the amplitude
of oscillation in time is smaller.) The two plots on
the right are for the relative values of $E_{\cal N}^{(0)}$ at spatial
separation $d$ to the value at infinite spatial separation, as given
in $(\ref{ENrel})$. In the upper-right plot, the brighter 
color corresponds to the higher value of $E_{{\cal N}{\rm rel}}$.}
\label{zeroS}
\end{figure}

Neglecting the mutual influences, the v-part of the zeroth-order
cross correlators read
\begin{eqnarray}
  \left<\right.Q_A(t), Q_B(t) \left.\right>_{\rm v}^{(0)} &=&
    {\hbar\over \pi \Omega^2 d}{\rm Re}{i\over \Omega+i \gamma}\left\{
      \left[ \Omega + e^{-2\gamma t} \left( \Omega +2\gamma
      e^{i\Omega t}\sin\Omega t\right)\right] {\cal S}_d \right. -
      \nonumber\\ && \left. e^{-\gamma t}
   \left[ \left(\Omega\cos\Omega t+\gamma \sin\Omega t\right)\left(
     {\cal S}_{d-t} + {\cal S}_{d+t}\right)
     +(\Omega+i\gamma)\sin\Omega t\left(
     {\cal C}_{d-t}-{\cal C}_{d+t}\right)
     \right]\right\},\label{QLQRv0}\\
  \left<\right.P_A(t), P_B(t) \left.\right>_{\rm v}^{(0)} &=&
    {\hbar\over \pi \Omega^2 d}{\rm Re}\,\, i (\Omega+i \gamma)\left\{
      \left[ \Omega + e^{-2\gamma t} \left( \Omega -2\gamma
      e^{i\Omega t}\sin\Omega t\right)\right] {\cal S}_d \right. -
      \nonumber\\ && \left. e^{-\gamma t}
   \left[ \left(\Omega\cos\Omega t-\gamma \sin\Omega t\right)\left(
     {\cal S}_{d-t}+{\cal S}_{d+t}\right)
     +(\Omega-i\gamma)\sin\Omega t\left( {\cal C}_{d-t}-{\cal C}_{d+t}
     \right)\right]\right\}, \label{PLPRv0} \\
  \left<\right.P_A(t), Q_B(t) \left.\right>_{\rm v}^{(0)} &=&
  \left<\right.Q_A(t), P_B(t) \left.\right>_{\rm v}^{(0)} \nonumber\\
    &=& {\hbar\gamma\over \pi \Omega^2 d}e^{-\gamma t}\sin\Omega t
    \,\,{\rm Re}\,\, \left\{ -2 e^{(-\gamma +i\Omega) t} {\cal S}_d +
     {\cal S}_{d-t} +{\cal S}_{d+t} + i \left( {\cal C}_{d-t} -
     {\cal C}_{d+t}\right)\right\}, \label{PLQRv0}
\end{eqnarray}
where
\begin{eqnarray}
{\cal S}_x &\equiv& {1\over 2}({\rm Ci}[(\Omega+ i\gamma)x]+
  {\rm Ci}[-(\Omega+ i\gamma)x])\sin[(\Omega+i\gamma)x]- 
  {\rm Si}[(\Omega+ i\gamma)x] \cos[(\Omega+i\gamma)x],\\
{\cal C}_x &\equiv& {1\over 2}({\rm Ci}[(\Omega+ i\gamma)x]+ 
  {\rm Ci}[-(\Omega+ i\gamma)x])\cos[(\Omega+i\gamma)x] + 
  {\rm Si}[(\Omega+ i\gamma)x]\sin[(\Omega+i\gamma)x],
\end{eqnarray}
with sine-integral Si$(x)={\rm si}(x)+ \pi/2$ and cosine-integral
Ci$(x)$ \cite{Arfken}.
The a-part of the zeroth-order correlators as well as the
two-point functions (for a single inertial detector),
$\left<\right.Q_j^2\left.\right>_{\rm v}^{(0)}$, $\left<\right.Q_j,
P_j \left.\right>_{\rm v}^{(0)}$, and $\left<\right.P_j^2
\left.\right>_{\rm v}^{(0)}$ 
are all independent of the spatial separation $d$ [for explicit
expressions see Eq. (25) in Ref. \cite{LCH08} and Appendix A in 
Ref. \cite{LH2006}].
So the $d$ dependence of the zeroth-order degrees of entanglement 
$E_{\cal N}^{(0)}$ and $\Sigma^{(0)}$ are all coming from 
$(\ref{QLQRv0})$-$(\ref{PLQRv0})$, which are due to the phase
difference of vacuum fluctuations that the detectors experience
locally.

Note that when
\begin{equation}
  d \to d_{\min} \equiv {1\over\Omega}e^{1-\gamma_e -\Lambda_1},
  \label{dmin}
\end{equation}
where $\gamma_e$ is the Euler constant and $\Lambda_1 \equiv -\ln 
\Omega\Delta t -\gamma_e$ corresponds to the time-resolution $\Delta t$
of our detector theory \cite{LH2006},
one has $\left<\right.{\cal R}_A(t), {\cal R}_B(t) \left.
\right>_{\rm v}^{(0)} \to \left<\right.{\cal R}_A(t)^2\left.
\right>_{\rm v}^{(0)}=\left<\right. {\cal R}_B(t)^2\left. 
\right>_{\rm v}^{(0)}$, ${\cal R}=P,Q$.
That is, the two detectors should be seen as
located at the same spatial point when $d\approx d_{\min}$ in our model,
which is actually a coarse-grained effective theory.
Let us call $d_{min}$ the ``merge distance."

\subsection{Early-time entanglement dynamics inside the light cone ($d<t$)}
\label{Earlyd<t}

In the weak-coupling limit ($\gamma\Lambda_1 \ll\Omega$),
when the separation $d$ is not too small,
the effect from the mutual influences comes weakly and slowly,
so the zeroth-order correlators dominate the early-time behavior of
the detectors. The asymptotic expansions of sine-integral and
cosine-integral functions read \cite{Arfken}
\begin{eqnarray}
  {\rm Ci}[(\Omega+i\gamma)x] &\approx& {i\pi\over 2}
     \left({x\over |x|}-1\right) +
     {\sin (\Omega+i\gamma)x\over(\Omega+i\gamma)x}, \label{Cixgg1}\\
  {\rm Si}[(\Omega+i\gamma)x] &\approx& {\pi\over 2}{x\over |x|}-
    {\cos(\Omega+i\gamma)x\over (\Omega+i\gamma)x},\label{Sixgg1}
\end{eqnarray}
for $\Omega, \gamma>0$, and $|(\Omega+ i\gamma)x| \gg 1$.
So in the weak-coupling limit,
from $t-d=0$ up to $t-d \sim O(1/\gamma)$, one has
\begin{equation}
  \left<\right.Q_A(t), Q_B(t) \left.\right>_{\rm v}^{(0)} \approx
    \theta(t-d){\sin\Omega d\over \Omega d}{\hbar\over 2\Omega}
    e^{-\gamma d} \left[ 1-e^{-2\gamma(t-d)}\right],
  \label{QLQRv0wcl}
\end{equation}
$\left<\right.P_A(t), P_B(t) \left.\right>_{\rm v}^{(0)}\approx\Omega^2
\left<\right.Q_A(t), Q_B(t) \left.\right>_{\rm v}^{(0)}$ and
$\left<\right.P_A(t), Q_B(t) \left.\right>_{\rm v}^{(0)}, \left<\right.
Q_A(t), P_B(t) \left.\right>_{\rm v}^{(0)} \sim O(\gamma/\Omega)$.
The $\theta(t-d)$ implies the onset of a clear interference pattern
($\sim \sin\Omega d/\Omega d$) inside the light cone, as shown in Fig.
\ref{zeroS}. This is mainly due to the sign flipping of the sine-integral 
function ${\rm Si}_{d-t}$ in
$(\ref{QLQRv0})$-$(\ref{PLQRv0})$ around $d=t$ when $d-t$ changes
sign. The $\theta(t-d)$ acts like that each detector starts to
``know" the existence of the other detector when they enter the
light cone of each other, though the mutual influences are not
considered here. In the next subsection we will see that there exists
some interference pattern of $O(\gamma)$ in $\Sigma$ even for $d>t$,
where no classical signal can reach one detector from the other.

\subsection{Outside the light cone ($d > t$)}
\label{outLC}

Before the first mutual influences from one detector reaches the other,
the zeroth-order results are exact. From $(\ref{Cixgg1})$ and
$(\ref{Sixgg1})$, when $d>t$ and $|\Omega+i\gamma|(d-t)\gg 1$, one has
\begin{eqnarray}
  \left<\right.Q_A(t), Q_B(t) \left.\right>_{\rm v}^{(0)} &\approx&
    {2\gamma\over\pi \Omega_r^4 d^2}\left[ 1
    + e^{-2\gamma t}\left(\cos\Omega t +{\gamma\over\Omega}\sin\Omega t\right)^2
    - {2d^2 e^{-\gamma t}\over d^2-t^2}
      \left(\cos\Omega t +{\gamma\over\Omega}\sin\Omega t\right)\right],
  \label{QAQBout}\\
  \left<\right.P_A(t), P_B(t) \left.\right>_{\rm v}^{(0)} &\approx&
    {2\gamma\over\pi d^2} e^{-2\gamma t} {\sin^2\Omega t\over\Omega^2} ,\\
  \left<\right.P_A(t), Q_B(t) \left.\right>_{\rm v}^{(0)} &=&
  \left<\right.Q_A(t), P_B(t) \left.\right>_{\rm v}^{(0)} \approx
    {2\gamma e^{-\gamma t} \over\pi \Omega_r^2 d^2}
    {\sin\Omega t\over\Omega} \left[ - e^{-\gamma t}\left(\cos\Omega t
      +{\gamma\over\Omega}\sin\Omega t\right) + {d^2\over d^2-t^2} \right],
  \label{QAPBout}
\end{eqnarray}
which makes the values of $E_{\cal N}$ and $\Sigma$ depend on $d$ and $t$;
that is, the dependence of the degree of entanglement on the spatial 
separation $d$ between the two detectors varies in time $t$, even before
they have causal contact with each other.

In the weak-coupling limit, with the initial state $(\ref{initGauss})$ and
$\Omega \gg \gamma\Lambda_j > \gamma$, $j=0,1$, one has
\begin{eqnarray}
  E_{{\cal N}rel} &\equiv& -\log_2 2c_-(t,d) -\left[ -
    \log_2 2c_-(t,\infty)\right] \nonumber\\
    &\approx& {\gamma \hbar \over \pi \ln 2}{{\cal X}\over |{\cal X}|}
    \sum_{n=0}^2 {a^\gamma_n\over b_\gamma} \cos n\Omega t +
    O(\gamma^2\Lambda_0, \gamma^2\Lambda_1)
\label{ENrel}
\end{eqnarray}
when $d>t$ and $|\Omega+i\gamma|(d-t)\gg 1$, where
\begin{eqnarray}
  a_0^\gamma &=& d^{-2}\left\{ \hbar^2\beta^2 + \alpha^2\left(
    -\alpha^2\beta^2\Omega^2+\beta^4+4\beta^2\hbar\Omega-\hbar^2\Omega^2\right)
    + |{\cal X}|(\alpha^2\Omega^2-\beta^2)+\right. \nonumber\\
    & & \,\,\,\,\,\,\,\left. 2\beta^2 e^{-2\gamma t}\left[\hbar^2+
    \alpha^2(\beta^2-2\hbar\Omega)-|{\cal X}|\right]\right\}e^{-2\gamma t},\\
  a_1^\gamma &=& -4(d^2-t^2)^{-1}\beta^2\left\{ 2\alpha^2\hbar\Omega + \left[
    \hbar^2+\alpha^2(\beta^2-2\hbar\Omega)-|{\cal X}|\right] e^{-2\gamma t}
      \right\}e^{-\gamma t},\\
  a_2^\gamma &=& d^{-2}\left\{ 4\hbar\Omega\alpha^2\beta^2 + \left[\beta^2\hbar^2
    + \alpha^2\left(\alpha^2\beta^2\Omega^2+\beta^4-4\beta^2\hbar\Omega+
    \hbar^2\Omega^2\right)-|{\cal X}|(\alpha^2\Omega^2+\beta^2)\right]
    e^{-2\gamma t}\right\},\\
  b_\gamma &=& \Omega^3\left\{ 2\hbar^2\Omega\alpha^2\beta^2 + \hbar\left[
    \alpha^2(\alpha^2\beta^2\Omega^2+\beta^4-4\beta^2\hbar\Omega) +
    (\alpha^2\Omega^2+\beta^2)(\hbar^2-|{\cal X}|)\right]e^{-2\gamma t}
    \right.\nonumber\\ & & \,\,\,\,\,\,\, \left.+
    (\hbar-\alpha^2\Omega)(\hbar\Omega-\beta^2)(\alpha^2\beta^2+\hbar^2
    -|{\cal X}|)e^{-4\gamma t}\right\},
\end{eqnarray}
with ${\cal X}\equiv \hbar^2-\alpha^2\beta^2$. So for ${\cal X}\not=0$
the relative degree of entanglement at separation $d$ to those for the
detectors at the same moment but separated at infinite distance
oscillates in frequency $\Omega$ and/or $2\Omega$, depending on the values
of $a_n^\gamma$. 
This explains the $(\cos \Omega t)/(d^2-t^2)$ pattern outside the
light cone in the upper-right plot of Fig. \ref{zeroS} and the small
oscillations before $t\approx 7.5$ in the lower-right plot of the same
figure, where $(\alpha, \beta)=(1.1, 4.5)$ so $(a_0^\gamma, a_1^\gamma,
a_2^\gamma)/b^\gamma \approx(1.94/d^2, -2.89/(d^2-t^2), 0.95/d^2)$ at early times.
Another example is, when $(\alpha, \beta)=(1.5, 0.2)$, one has $(a_0^\gamma,
a_1^\gamma, a_2^\gamma)/b^\gamma \approx (-4.68/d^2, -0.06/(d^2-t^2),
4.74/d^2)$ at early times, so the $d^{-2}\cos 2\Omega t$ pattern dominates
at large $d$ in the bottom-right plot of Fig. \ref{firstS2}.
For these cases, the larger the separation, the weaker the entanglement (in
terms of the logarithmic negativity) at some moments, but the stronger the
entanglement at other moments.

The sudden switching on of interaction at $t=0$ in our model will create
additional oscillation patterns outside the light cone. However, as shown in
$(\ref{ENrel})$, those oscillations are suppressed in the weak-coupling limit by
$O(\gamma \Lambda_0)$ of the above results. Here $\Lambda_0\equiv -
\ln\Omega\Delta t_0 -\gamma_e$ with $\Delta t_0$ corresponds to the time
scale of switching on the coupling between the detectors and the quantum field
(see Sec.IIIB in Ref.\cite{LH2006} for details).

When $\beta^2 = \hbar^2/\alpha^2$ or ${\cal X}=0$, the detectors are initially
separable and
\begin{eqnarray}
  \Sigma &\approx& {\hbar^2\over 16\alpha^4\pi^2 \Omega^4}\left\{ \pi\Omega
   (\hbar-\alpha^2\Omega)^2 e^{-2\gamma t}(1-e^{-2\gamma t}) +
   \right.\nonumber\\ & & \left.
   2\gamma\Lambda_1\left[ 2\hbar\alpha^2\Omega(1-e^{-2\gamma t})
   + h^2 e^{-2\gamma t}(1-\cos 2\Omega t)+ \alpha^4\Omega^2e^{-2\gamma t}
   (1+\cos 2\Omega t)\right] \right\}^2 + O(\gamma^2),
\end{eqnarray}
outside the light cone, which is always positive so the detectors are always
separable. When we increase the coupling strength $\gamma$, we find that
the values of $\Sigma$ are pushed further away from those negative values of
entangled states. In Appendix \ref{EarlyAna} we also see that quantum
entanglement is only created deep in the light cone. Therefore in our model
we see no evidence of entanglement generation outside the light cone.

For $|{\cal X}| \not=0$ but sufficiently small, the detectors are initially
entangled, but after a very short-time scale $O(e^{-\gamma_e-(\Lambda_0/2)})$
the value of $\Sigma$ jumps to $(\hbar^2\gamma\Lambda_1 \alpha/\beta\pi)^2
-(\hbar{\cal X}/4\alpha\beta)^2$, which could be positive so that the
detectors become separable. In these cases quantum entanglement could revive
later as $\Sigma$ is oscillating with an amplitude proportional to $\gamma
\Lambda_1$, while these revivals of entanglement do not last more than a few
periods of the intrinsic oscillation in the detectors.

\subsection{Breakdown of the zeroth-order results}
\label{0bad}

At late times $t\gg \gamma^{-1}$, all $\left<\right. .. \left.
\right>_{\rm a}$ vanish, so $\left<\right. .. \left.\right>_{\rm v}$
dominate and the nonvanishing two-point correlation functions read
\begin{eqnarray}
  \left<\right.Q_A, Q_B \left.\right>^{(0)}|_{t\gg \gamma^{-1}} &\approx&
  {\hbar\over \pi \Omega d}{\rm Re}{i{\cal S}_d \over \Omega+i \gamma},
    \label{QAQB0late}\\
  \left<\right.P_A, P_B \left.\right>^{(0)}|_{t\gg \gamma^{-1}} &\approx&
    {\hbar\over \pi \Omega d}{\rm Re} (i\Omega -\gamma) {\cal S}_d ,
    \label{PAPB0late}\\
  \left<\right.Q_A^2\left.\right>^{(0)}|_{t\gg \gamma^{-1}} =
    \left<\right.Q_B^2\left.\right>^{(0)}|_{t\gg \gamma^{-1}} &\approx&
    {i\hbar\over 2\pi\Omega}\ln {\gamma-i\Omega\over \gamma+i\Omega},\\
  \left<\right.P_A^2\left.\right>^{(0)}|_{t\gg \gamma^{-1}} =
    \left<\right.P_B^2\left.\right>^{(0)}|_{t\gg \gamma^{-1}} &\approx&
    {\hbar\over\pi} \left\{ {i \over 2\Omega} (\Omega^2-\gamma^2)\ln
    {\gamma-i\Omega\over \gamma+i\Omega} + \gamma\left[ 2\Lambda_1 -
    \ln \left( 1+{\gamma^2\over\Omega^2}\right)\right]\right\},
\end{eqnarray}
from $(\ref{QLQRv0})$-$(\ref{PLQRv0})$ and from Ref. \cite{LH2006}.

\begin{figure}
\includegraphics[width=8cm]{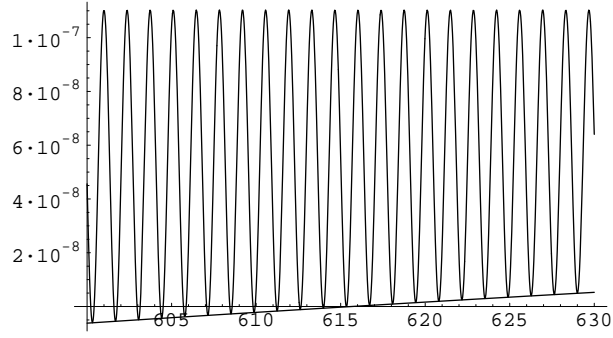}
\caption{The oscillating curve represents the value of
$\Upsilon^{(0)}$ (defined in $(\ref{0thUnc})$) as a function of $d$.
The bottom curve represents its lower bound (Eq.$(\ref{minU})$).
It becomes negative when $d < 616$, which signifies the violation
of uncertainty relation. To rectify this, one needs to add on the
mutual influences, as shown in Fig.  \ref{dent}.
Here $\gamma=10^{-4}$, $\Omega=2.3$, $\Lambda_0=\Lambda_1=25$.}
\label{zerounc}
\end{figure}

When $d \to \infty$, the cross correlators vanish and the uncertainty
relation reads
\begin{equation}
  \Upsilon^{(0)}|_{t\gg \gamma^{-1}} \equiv \det\left[
  {\bf V}^{(0)}|_{t\gg \gamma^{-1}}+ {i\over 2}\hbar {\bf M}\right]
  \approx \left( \left<\right.Q_A^2\left.\right>^{(0)}
    \left<\right.P_A^2\left.\right>^{(0)}|_{t\gg \gamma^{-1}} -
    {\hbar^2\over 4}\right)^2 \ge 0, \label{0thUnc}
\end{equation}
for sufficiently large $\Lambda_1$ \cite{LH2006}, so the uncertainty
relation holds perfectly. However, observing that $\left|{\cal S}_d
\right|\approx\pi e^{-\gamma d}$  for $d$ large enough but still finite,
the late-time $\Upsilon^{(0)}$ can reach the lowest values:
\begin{eqnarray}
  && \left( \left<\right.Q_A^2(t)\left.\right>^{(0)} \left<\right.
  P_A^2(t)\left.\right>^{(0)}|_{t\gg \gamma^{-1}} - {\hbar^2\over 4}
  \right)^2 + {\hbar^4 e^{-4\gamma d} \over 16 \Omega_r^4 d^4}
  - \nonumber\\ & & {\hbar^2 e^{-2\gamma d} \over 4d^2}\left[
  {\hbar^2\over 2 \Omega_r^2} + \left(\left<\right.Q_A^2(t)
  \left.\right>^{(0)}|_{t\gg \gamma^{-1}}\right)^2 + \Omega_r^{-4}
  \left(\left<\right.P_A^2(t)\left.\right>^{(0)}|_{t\gg \gamma^{-1}}
  \right)^2\right]. \label{minU}
\end{eqnarray}
This zeroth-order result suggests that the uncertainty relation can
fail if $d$ is not large enough to make the value of the second line 
of $(\ref{minU})$ overwhelmed by the first line [see Fig. \ref{zerounc}].
When this happens the zeroth-order results break down. Therefore
to describe the long-time entanglement dynamics
at short distances $d$ the higher-order corrections from the mutual
influences must be included for consistency.

When $\gamma \ll \gamma\Lambda_1 \ll \Omega$, one has a simple estimate
that the late-time $\Upsilon^{(0)}$ becomes negative if $d$ is smaller
than about $d_{0}\approx \pi /2\Lambda_1\gamma$, which is much
greater than $d_{ins}$ found in Sec.\ref{instab}.

\section{Entanglement at late times}
\label{resient}

Since all $q_i^{(j)}$ vanish at late times in the stable regime (see 
Appendix A), the late-time correlators consist of $q_j^{(\pm)}$ only,
for example,
\begin{equation}
  \left<\right. Q_B^2 \left.\right>|_{t\to \infty} =\int {\hbar d^3 k
    \over(2\pi)^3 2\omega} q_B^{(+)}(t,{\bf k}) q_B^{(-)}(t,{\bf k})
    |_{t\to \infty},
\end{equation}
where $q_B^{(+)}(t,{\bf k})|_{t\to \infty}$ is given by $(\ref{lateTqp})$
and $q_B^{(-)}(t,{\bf k})|_{t\to \infty}$ is its complex conjugate.
After some algebra, we find that the value of the nonvanishing 
correlators at late times can be written as
\begin{eqnarray}
  \left<\right. Q_A^2 \left.\right>|_{t\to \infty} =
    \left<\right. Q_B^2 \left.\right>|_{t\to \infty} &=&
    2 {\rm Re}\left( {\cal F}_{0+} + {\cal F}_{0-} \right), \label{LTQ2}\\
  \left<\right. Q_A, Q_B \left.\right>|_{t\to \infty} &=&
    2 {\rm Re}\left( {\cal F}_{0+} - {\cal F}_{0-} \right), \\
  \left<\right. P_A^2 \left.\right>|_{t\to \infty} =
    \left<\right. P_B^2 \left.\right>|_{t\to \infty} &=&
    2 {\rm Re}\left( {\cal F}_{2+} + {\cal F}_{2-} \right), \\
  \left<\right. P_A, P_B \left.\right>|_{t\to \infty} &=&
    2 {\rm Re}\left( {\cal F}_{2+} - {\cal F}_{2-} \right), \label{LTPx}
\end{eqnarray}
where
\begin{equation}
  {\cal F}_{c\pm}(\gamma,\Omega, d) \equiv {\hbar i\over 4\pi}
    \int_0^{\omega_{max}}d\omega {\omega^c\over \omega^2+2i\gamma\omega- 
      \Omega_r^2\pm {2\gamma\over d}e^{i\omega d} },
\end{equation}
and $\omega_{max}$ is the high frequency (UV) cutoff corresponding to
$\Lambda_1$.

In the stable regime one can write ${\cal F}_{c\pm}$ 
in a series form:
\begin{eqnarray}
  {\cal F}_{c\pm}(\gamma,\Omega, d) &=& {\hbar i\over 4\pi}
    \int_0^{\omega_{max}}d\omega
      {\omega^c\over \omega^2+2i\gamma\omega-\Omega^2-\gamma^2}
      \sum_{n=0}^\infty \left[{\mp {2\gamma\over d}e^{i\omega d} \over
      \omega^2+2i\gamma\omega-\Omega^2 - \gamma^2}\right]^n \nonumber\\
  &=& {\hbar i\over 4\pi} \int_0^{\omega_{max}} d\omega \sum_{n=0}^\infty
    {1\over n!}\left[ \mp {\gamma\over \Omega d}e^{i\omega d}\partial_\Omega
     \right]^n {\omega^c\over \omega^2+2i\gamma\omega-\Omega^2-\gamma^2},
  \label{Fseries}
\end{eqnarray}
so we have
\begin{eqnarray}
  {\cal F}_{0\pm}(\gamma,\Omega, d) &=& {\hbar \over 4\pi}\left\{
    {i\over 2\Omega}\ln {\gamma-i\Omega\over\gamma+i\Omega} +
    \sum_{n=1}^\infty {1\over n!} \left[ \mp{\gamma\over \Omega d}
    \partial_\Omega\right]^n {\rm Re}\, {i\over\Omega}
    e^{(\gamma+i\Omega)nd}\Gamma[0, (\gamma+i\Omega)nd]\right\},\\
  {\cal F}_{2\pm}(\gamma,\Omega, d) &=& {\hbar \over 4\pi}\left\{
    {i\over 2\Omega}\left(\Omega^2-\gamma^2\right)
      \ln {\gamma-i\Omega\over\gamma+i\Omega}+\gamma\left[2\Lambda_1 -
      \ln\left( 1+{\gamma^2\over\Omega^2}\right)\right] + \right.
    \nonumber\\ & & \left. \sum_{n=1}^\infty {1\over n!}
    \left[ \mp{\gamma\over \Omega d}
    \partial_\Omega\right]^n {\rm Re} {i\over\Omega}e^{(\gamma+i\Omega)nd}
     (\gamma+i\Omega)^2\Gamma[0, (\gamma+i\Omega)nd]  \right\},
\end{eqnarray}
for large frequency cutoff $\omega_{max}$, or the corresponding
$\Lambda_1$.

Substituting the late-time correlators $(\ref{LTQ2})$-$(\ref{LTPx})$
into the covariance matrix ${\bf V}$, we get
\begin{eqnarray}
  \Sigma |_{t\to \infty} &=&
  \left( 16{\rm Re}{\cal F}_{0+}{\rm Re}{\cal F}_{2-}
    -{\hbar^2\over 4}\right)
  \left( 16{\rm Re}{\cal F}_{0-}{\rm Re}{\cal F}_{2+}
    -{\hbar^2\over 4}\right),\\
  \Upsilon|_{t\to\infty} &=&
  \left( 16{\rm Re}{\cal F}_{0+}{\rm Re}{\cal F}_{2+}
    -{\hbar^2\over 4}\right)
  \left( 16{\rm Re}{\cal F}_{0-}{\rm Re}{\cal F}_{2-}
    -{\hbar^2\over 4}\right).
  \label{lateUnc}
\end{eqnarray}
Numerically we found that $16{\rm Re}{\cal F}_{0+}{\rm Re}{\cal F}_{2-}
-(\hbar^2/4)$ and $\Upsilon|_{t\to\infty}$ are positive definite in
the cases considered in this paper. We then identify the late-time
symplectic spectrum $(c_+, c_-)|_{t\to\infty}= 
(4\sqrt{{\rm Re}{\cal F}_{0+}{\rm Re}{\cal F}_{2-}}, 4\sqrt{{\rm Re}
{\cal F}_{0-}{\rm Re}{\cal F}_{2+}})$. So if $16{\rm Re}
{\cal F}_{0+}{\rm Re}{\cal F}_{2-}-(\hbar^2/4)$ is negative, then 
$\Sigma < 0$, $E_{\cal N}>0$, and the detectors are entangled.

\begin{figure}
\includegraphics[width=8cm]{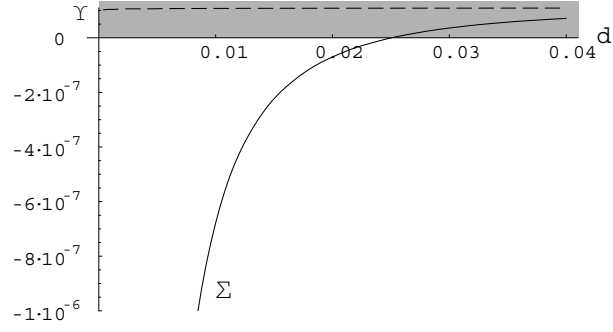}
\caption{Plots for $\Sigma$ (solid curve) and $\Upsilon$ (dashed curve)
at late times as a function of $d$, with parameters the same as those
in Fig.  \ref{zerounc}. Two detectors are separable when $\Sigma\ge 0$
(shaded zone). One can see that $\Sigma$ becomes negative when
$d < 0.025$. With the mutual influences included, the uncertainty
relation [see Eq.$(\ref{Uncert})$ and below] now holds for all $d$.}
\label{dent}
\end{figure}

In the weak-coupling limit, keeping the correlators to $O(\gamma/d)$,
we have
\begin{eqnarray}
16{\rm Re}{\cal F}_{0+}{\rm Re}{\cal F}_{2-} -{\hbar^2\over 4} &\approx&
  {\hbar^2\gamma\Lambda_1\over\pi\Omega} -
  {\hbar^2\over\Omega^3} {\rm Re}\left\{ \left[
  {i\gamma\Omega\over \pi d} + {2\gamma^2\Lambda_1\over\pi^2 d}(i+\Omega d)
  \right] e^{i\Omega d}\Gamma[0,i\Omega d]\right\}, \label{soldent}
\end{eqnarray}
which is positive as $d\to\infty$, but negative when $d \to 0_+$. So
$(\ref{soldent})$ must cross zero at a finite ``entanglement distance"
$d_{ent} > 0$, where $\Sigma =0$. For $d < d_{ent}$, the detectors will have
residual entanglement, while for $d>d_{ent}$, the detectors are separable at
late times.

For small $\gamma$, $d_{ent}$ is almost independent of $\gamma$. We find
that when $\gamma\Lambda_1\ll\Omega$ and $\Lambda_1 \gg 1$,
\begin{equation}
  d_{ent} \approx {\pi/2\Omega\over \Lambda_1-\ln{\pi\over 2\Lambda_1}}.
  \label{dentdef}
\end{equation} 
will be a good estimate if $d_{ent}\ll 1$. 
Here $d_{ent}$ is still much larger than the
``merge distance" $d_{min}$ in $(\ref{dmin})$. For example, as shown
in Fig.  \ref{dent}, when $\gamma = 0.0001$, $\Omega = 2.3$,
$\Lambda_1 = 25$, one has $d_{ent}\approx 0.025$, which is quite a
bit greater than the ``radius of instability"
$2\gamma/\Omega_r^2 \approx 3.8\times 10^{-5}$, and much greater
than the merge distance $d_{min}\approx 9 \times 10^{-12}$.

A corollary follows. If the initial state of the two detectors with
$d< d_{ent}$ is separable, then the residual entanglement implies
that there is an entanglement creation during the evolution. In
contrast, if the initial state of the two detectors with $d >
d_{ent}$ is entangled, then the late-time separability implies that
they disentangled in a finite time. Examples will be given in the
next section.

Note that the ill behavior of $\Upsilon^{(0)}$ has been cured by mutual
influences. The uncertainty function
$(\ref{lateUnc})$ is positive for all $d$ at late times.

Note also that, while the corrections from the mutual influences to
$\left<\right. Q_A^2\left.\right>|_{t\to \infty}$ and $\left<\right.
P_A^2 \left.\right>|_{t\to \infty}$ are $O(\gamma/d)$, the mutual
influences have been included in the leading order 
approximation for the cross correlators.
Indeed, in $(\ref{Fseries})$, even as low as $n=1$, we have had
\begin{eqnarray}
  \left<\right. Q_A,Q_B \left.\right>|_{t\to \infty} &\approx&
  \left<\right. Q_A,Q_B \left.\right>^{(0)}|_{t\to \infty}-
    {2 \hbar \gamma\over \pi} {4\gamma\over d} \int_0^\infty d\omega {\omega
    \left[(\Omega^2_r-\omega^2)\cos\omega d-2\gamma\omega\sin\omega d\right]
    \over \left[ (\omega^2-\Omega^2_r)^2+4 \gamma^2\omega^2\right]^2}.
\label{1stgamma}
\end{eqnarray}
However, this is slightly different from the approximation with the 
first-order mutual influences included. Writing the $n=0$ and $n=1$ 
terms in Eq. $(\ref{qjp})$ as
\begin{equation}
   q_j^{(+)} \approx q_{j,n=0}^{(+)} + q_{j,n=1}^{(+)},
\end{equation}
then the approximated cross correlator with the first-order mutual
influences included is the $\omega$ integration of Re
$[(q_{A,n=0}^{(+)} + q_{A,n=1}^{(+)})(q_{B,n=0}^{(+)} +
q_{B,n=1}^{(+)})]$, but in $(\ref{1stgamma})$ only  Re
$[q_{A,n=0}^{(+)}q_{B,n=0}^{(+)}+ q_{A,n=0}^{(+)} q_{B,n=1}^{(+)}+
q_{A,n=1}^{(+)}q_{B,n=0}^{(+)}]$ contribute, though there are
$O(\gamma^0)$ terms in $q_{A,n=1}^{(+)}q_{B,n=1}^{(+)}$. The latter
is small for $\Omega d \gg 1$, and will be canceled by the mutual
influences of higher-orders.

\section{Entanglement dynamics in weak-coupling limit}

\subsection{Disentanglement at very large distance}

Suppose the two detectors are separated far enough $(d \gg \Omega)$ so
that the cross correlations and the mutual influences can be safely ignored.
Then in the weak-coupling limit ($\Omega \gg \gamma \Lambda_1$)
the zeroth-order results for the v-part of the self correlators dominate,
so that \cite{LCH08}
\begin{eqnarray}
  \left<\right.Q_A^2\left.\right>_{\rm v} = \left<\right.Q_B^2\left.
    \right>_{\rm v}&\approx& {\hbar\over 2\Omega} \left(1-e^{-2\gamma t}\right),
    \label{Q2weak0}\\
  \left<\right.P_A^2\left.\right>_{\rm v} = \left<\right.P_B^2\left.
    \right>_{\rm v}&\approx& {\hbar\over 2}\Omega \left(1-e^{-2\gamma t}\right) +
    {2\over\pi}\hbar\gamma\Lambda_1 , \label{P2weak0}
\end{eqnarray}
and $\left<\right.Q_A,P_A\left.\right>_{\rm v} = \left<\right.Q_B,P_B\left.
\right>_{\rm v} \sim O(\gamma)$, while the v-part of the cross correlators are
vanishingly small. This is exactly the case we have considered in Sec. IV A 2
of Ref. \cite{LCH08}, where we found
\begin{equation}
  \Sigma \approx {\hbar^2 e^{-4\gamma t}\over 16\alpha^2\beta^2\Omega^2}
  \left[ Z_8 \left( e^{-4\gamma t} -2 e^{-2\gamma t}\right) + Z_4\right]
  + {\hbar^3 \gamma\Lambda_1 \over 4\pi\alpha^2\beta^2\Omega^2}
    Z_2 e^{-2\gamma t} +
    {\hbar^4\over \pi^2\Omega^2} \gamma^2\Lambda_1^2,
\label{Sig0}
\end{equation}
with $Z_8\ge 0$, $Z_8-Z_4 \ge 0$ and $Z_2 \ge 0$ [$Z_8$, $Z_4$ and
$Z_2$ are parameters depending on $\alpha$ and $\beta$, defined in
Eqs.$(37)$, $(38)$ and $(41)$ of Ref. \cite{LCH08}, respectively.]
Accordingly the detectors always disentangle in a finite time. There
are two kinds of behaviors that $\Sigma$ could have. For $Z_4 >0$,
the disentanglement time is a function of $Z_4$, $Z_8$ and $\gamma$,
\begin{equation}
  t^{(0)}_{dE>} \approx -{1\over 2\gamma}\ln\left(
    1-\sqrt{1-{Z_4\over Z_8}}\right), \label{tdEZ4p}
\end{equation}
while for $Z_4<0$, the disentanglement time is much longer,
\begin{equation}
  t^{(0)}_{dE<} \approx {1\over 2\gamma}
    \ln { |Z_4|\pi/(2\hbar \gamma\Lambda_1)
  \over Z_2 + \sqrt{Z_2^2-4\alpha^2\beta^2 Z_4} } , \label{tdEZ4n}
\end{equation}
and depends on $\Lambda_1$.

\subsection{Disentanglement at large distance}

When $d$ is large (so $1/\Omega d$ is small) but not too large to make
all the mutual influences negligible,
while the zeroth-order results for the v-part of the self-correlators
$(\ref{Q2weak0})$ and $(\ref{P2weak0})$ are still good, the 
first-order correction [$n=1$ terms in ($\ref{qjp}$)] to the cross
correlators $\left<\right. Q_A, Q_B \left.\right>$ can be of the
same order of $\left<\right. Q_A, Q_B \left.\right>^{(0)}$ (a similar
observation on the late-time correlators has been mentioned in the
end of Sec. \ref{resient}). Including the first-order correction,
for $d > O(1/\sqrt{\gamma\Omega})$, we have a simple expression,
\begin{eqnarray}
  \left<\right. Q_A, Q_B \left.\right>_{\rm v} &=&
  \left<\right. Q_A, Q_B \left.\right>^{(0)}_{\rm v} + \theta(t-d)
  {\hbar\over 2\Omega}{\sin\Omega d\over\Omega d}e^{-\gamma d}
  \left[-1 + e^{-2\gamma (t-d)}\left(1+2 (t-d)\gamma \right)
+ O(\gamma/\Omega) \right] \nonumber\\
&\approx&  \theta(t-d){\hbar \over \Omega}{\sin\Omega d\over\Omega d}
  e^{-\gamma d}\gamma (t-d)e^{-2\gamma (t-d)} ,  \label{QLQRweak}
\end{eqnarray}
and $\left<\right. P_A, P_B \left.\right>_{\rm v}\approx \Omega^2
\left<\right. Q_A, Q_B \left.\right>_{\rm v}$ with other two-point functions
$\left<\right. .. \left.\right>_{\rm v}$ being $O(\gamma)$ for all $t$.
Here $\left<\right. Q_A, Q_B \left.\right>^{(0)}_{\rm v}$ in the weak-coupling
limit has been shown in $(\ref{QLQRv0wcl})$. The above approximation is good
over the time interval from $t=0$ up to $e^{-2\gamma(t-d)}>O(\gamma/\Omega)$,
namely, before $t-d \sim O(-\gamma^{-1}\ln(\gamma/\Omega))$.

\begin{figure}
\includegraphics[width=8cm]{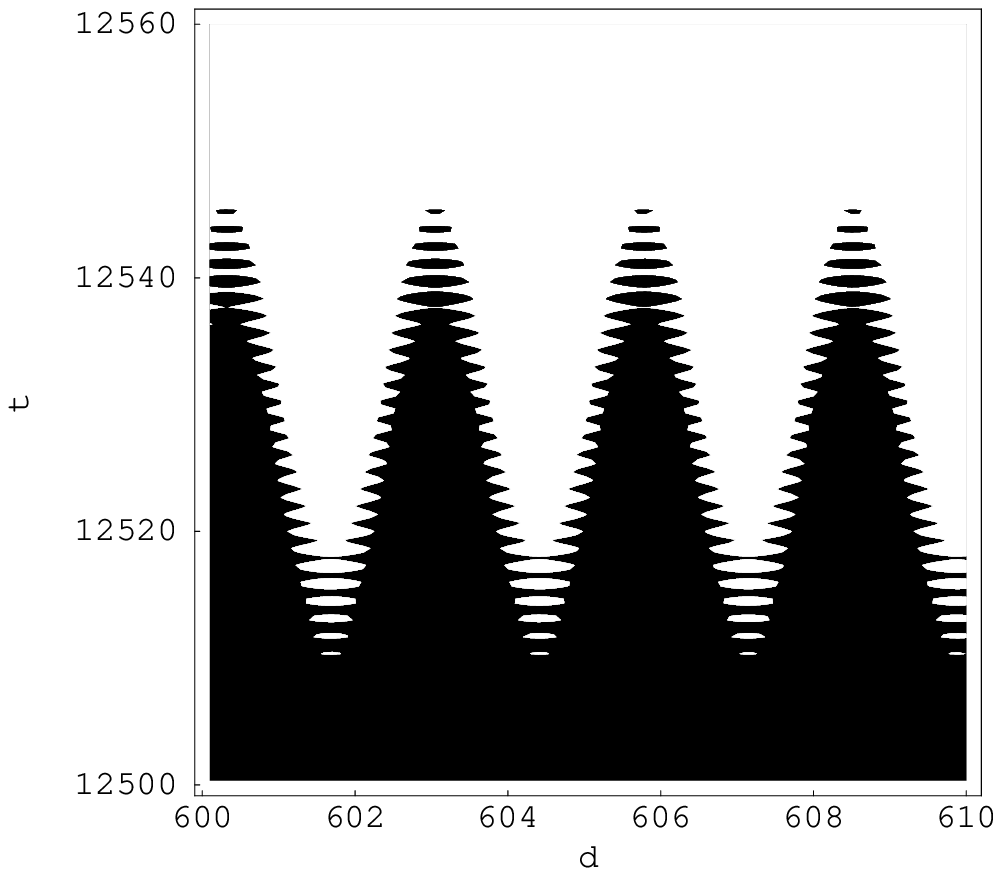}
\includegraphics[width=8cm]{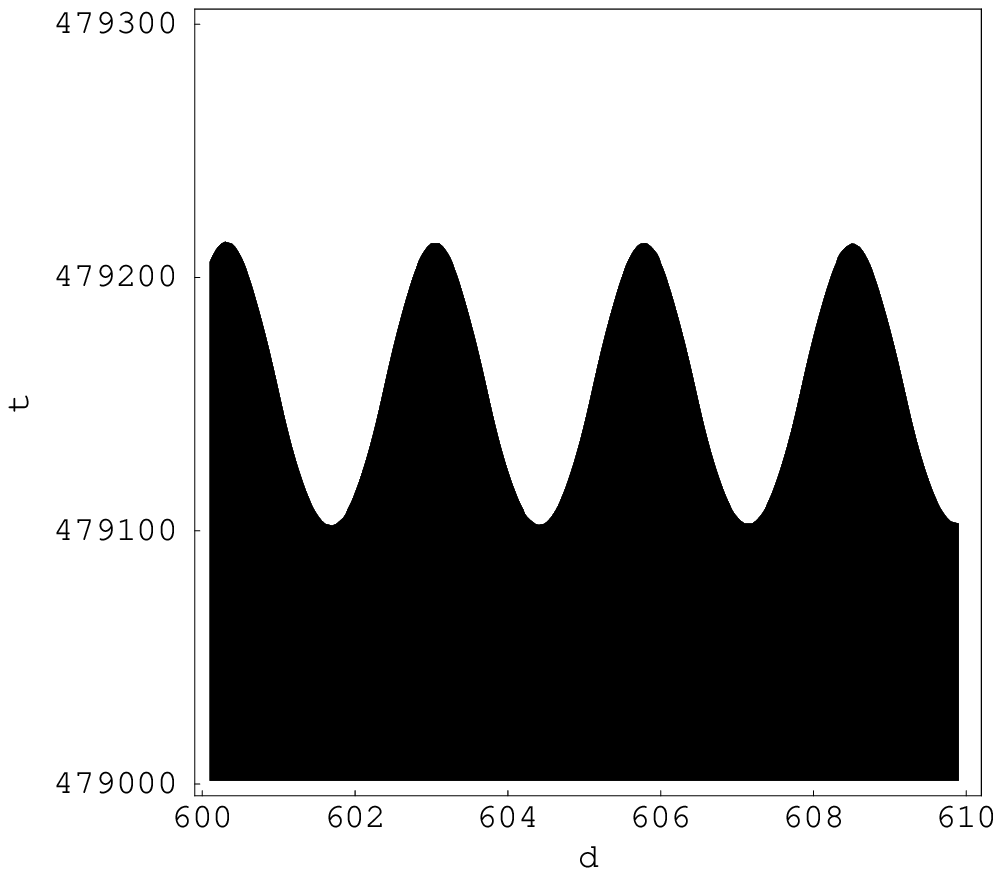} 
\caption{The plot of $\Sigma$  as a function of $d$ and $t$, 
up to the first-order correction. 
$\Sigma$ is negative in the dark region and positive in the bright region.
For a fixed $d$, the disentanglement time $t_{dE}$ is at the border
of the lowest dark region or the earliest time that the detectors becomes
separable. The interference pattern in Fig.  \ref{zeroS} for $\Sigma$
at early times signifies that the disentanglement time $t_{dE}$ is
longer or shorter than those at $d\to\infty$ [Eqs. ($\ref{tdEZ4p}$]
and ($\ref{tdEZ4n})$). The gridded profile in the left plot shows
that after $t_{dE}$ there could be some short-time revivals of
entanglement. Here the parameters are the same as those in Fig.
\ref{zeroS} except $(\alpha,\beta)=(1.5,0.2)$ in the left plot
and $(1.1,4.5)$ in the right (cf. Fig. 3 in Ref. \cite{LCH08}).}
\label{tdEvsd}
\end{figure}

Still, in this first-order approximation, $\left<\right. Q_A, Q_B
\left.\right>_{\rm v}$ and $\left<\right. P_A, P_B \left.\right>_{\rm
v}$ are the only correlators depending on the separation $d$.
Inserting those approximated expressions for the correlators into the
definition of $\Sigma$ or $E_{\cal N}$, we find that the interference 
pattern in $d$ for the relative values of $\Sigma$ or $E_{\cal N}$ 
at early times (Fig. \ref{zeroS}) can last through the disentanglement
process to make the disentanglement time $t_{dE}$ longer or shorter
than those at $d\to\infty$, though the contrast decays noticeably 
compared with those at early times. Two examples are shown in Fig.
\ref{tdEvsd}. For $Z_4>0$, the disentanglement time is about
\begin{equation}
  t_{dE>} \approx t^{(0)}_{dE>} - {Z_6
  \left(t^{(0)}_{dE>}-d\right)e^{\gamma d}\sin\Omega d \over Z_8 d
  \left(1- e^{-2\gamma t^{(0)}_{dE>}}\right) + Z_6 \left[1-2\gamma
  \left(t^{(0)}_{dE>}-d \right)\right]e^{\gamma d}\sin\Omega d},
\end{equation}
where $Z_6 \equiv (\hbar^2-\alpha^2\beta^2)(\alpha^2\Omega^2-\beta^2)$
[Fig.  \ref{tdEvsd} (left)].
In this case the disentanglement time can be short compared to the 
time scale $O(n/\gamma)$, $n\in N$ when the higher-order corrections 
$q_n$ from mutual influences reach their maximum values (see Sec.
\ref{EvoOp}). So in the weak-coupling limit the above estimate could be
good from large $d$ all the way down to $\Omega d \sim O(1)$ but still
much greater than $\Omega d_{ent}$. If this is true, the difference of
disentanglement times for different spatial separations can be significant
at small $d$. For example, for $(\alpha, \beta) =(1.5, 0.2)$ with other
parameters the same as those in Fig. \ref{tdEvsd}, the disentanglement
time at $d\approx 4.4934/\Omega$ (where $\sin\Omega d/ \Omega d$
is the global minimum) is over $1.6$ times longer than those for
$d\approx 7.7253/\Omega$ (where the first peak of $\sin\Omega d/ 
\Omega d$ is located). 

For $Z_4<0$, the correction of $\sin \Omega d$ is below the precision of
$t^{(0)}_{dE<}$ estimated in $(\ref{tdEZ4n})$. Here we just show the
numerical result up to the first-order mutual influences
in Fig.  \ref{tdEvsd} (right), which shows that the interference pattern
in $d$ is suppressed but still nonvanishing for large disentanglement times.

\subsection{Entanglement generation at very short distance}
\label{createEnt}

\begin{figure}
\includegraphics[width=8cm]{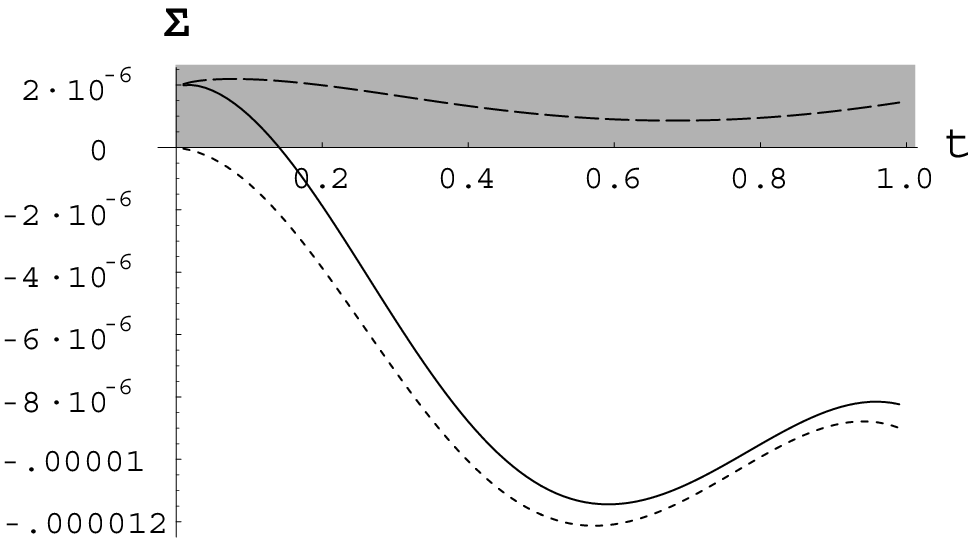} 
\includegraphics[width=8cm]{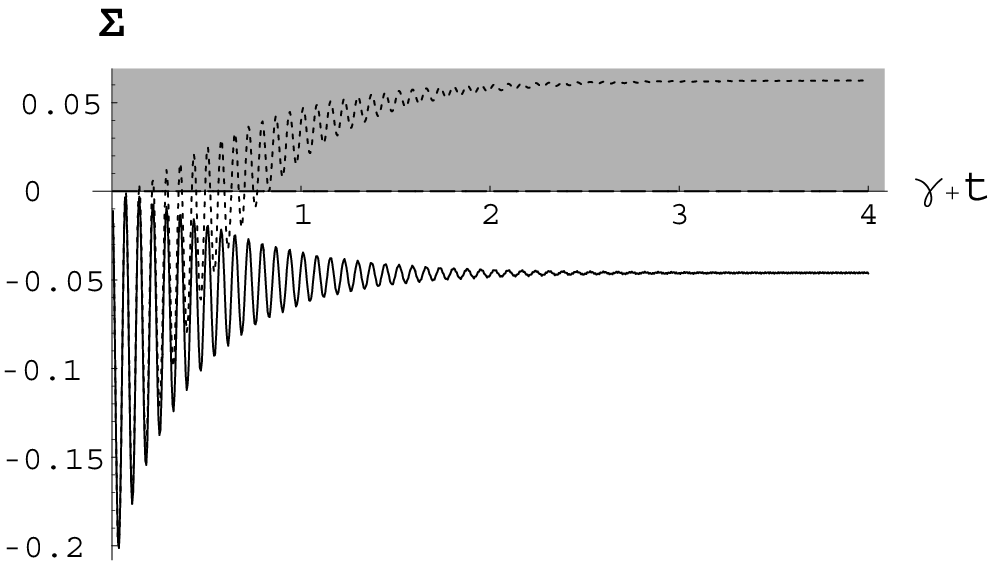}\\ 
\includegraphics[width=8cm]{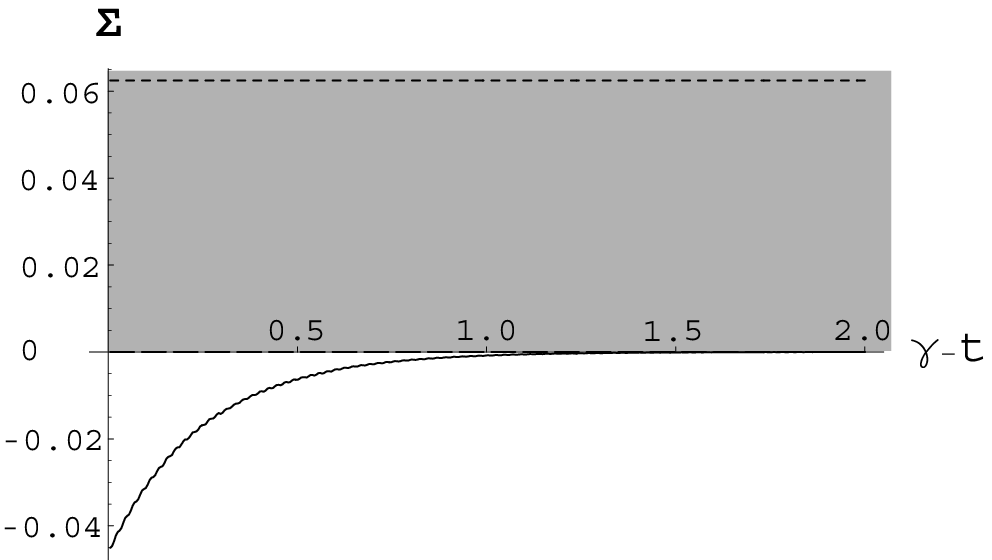} 
\includegraphics[width=8cm]{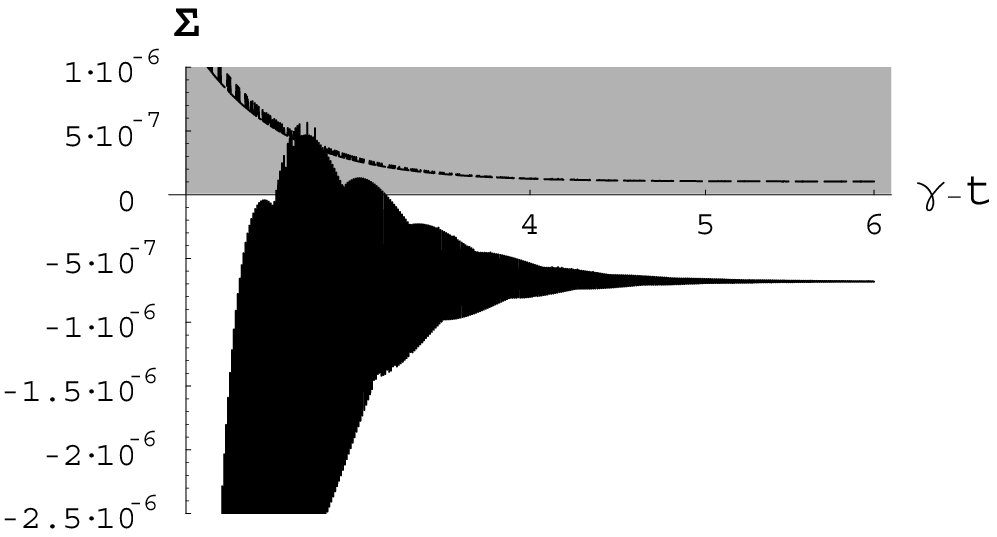} 
\caption{(Upper left)
The solid curve and the long-dashed curve represent the values of $\Sigma$
and $\Upsilon$, respectively, while the dotted line is for the value of
$\Sigma$ with all $\left<\right. .. \left.\right>_{\rm v}$ set to zero.
The detectors are separable initially (the parameters here are the same as
those in Figs. \ref{zerounc} and \ref{dent} except $d=0.01$ and
$(\alpha,\beta)=(1,1)$). Quantum entanglement has been generated after
$t \approx 0.15$, and $\Sigma$ oscillates in frequency $\Omega_+$ at early
times. Around the time scale $t\sim 1/\gamma_+$ ($\approx 5000$ here)
(upper right), $\Sigma$ has an oscillation with long period $\pi/(\Omega_-
- \Omega_+)$ $(\approx 361.4)$. $\Sigma$ appears to be settling down at a
value (about $-0.046$ here) depending in this case only on the value
$\alpha$ in the initial data. However, in a much longer time scale
$t\sim 1/\gamma_-$ ($\approx 1.13\times 10^8$) (lower left), one sees
that the value of $|\Sigma|$ is actually decaying exponentially to the
late-time value ($\approx -6.8\times 10^{-7}$) consistent with the
results in Fig.  \ref{dent} with $d=0.01$ and independent of the initial
data of the detectors.}
\label{EntGen}
\end{figure}

When $\Omega d \sim O(\epsilon)$, $\gamma/\Omega \sim O(\epsilon^2)$,
and $\epsilon \ll 1$, one can perform a dimensional reduction on
the third derivatives in $(\ref{eomsmalld})$, namely,
\begin{equation}
   \dddot{q}_\pm^{(j)} \approx -{\Omega_r^2 \mp {2\gamma\over d} \over
      1\mp\gamma d}\dot{q}_\pm^{(j)},
\end{equation}
to obtain, up to $O(\epsilon^5)$,
\begin{eqnarray}
  \ddot{q}^{(j)}_\pm +2\gamma_\pm \dot{q}^{(j)}_\pm  +\Omega^2_\pm
    \dot{q}^{(j)}_\pm &\approx& 0, \label{eomqje5}\\
  \ddot{q}^{(+)}_\pm +2\gamma_\pm \dot{q}^{(+)}_\pm  +\Omega^2_\pm
    \dot{q}^{(+)}_\pm &\approx& \lambda_\pm \left( e^{-i k_1 d/2 }
    \pm e^{i k_1 d/2}\right)e^{-i\omega t}, \label{eomqpme5}
\end{eqnarray}
where $j=A,B$, $q_\pm^{(+)} \equiv q_A^{(+)} \pm q_B^{(+)}$ and
\begin{eqnarray}
  \gamma_- &\equiv& {\gamma d^2\over 6}{\left(\Omega_r^2 +
    {2\gamma\over d}\right)\over (1+\gamma d)^2}, \\
  \gamma_+ &\equiv& {2\gamma\over 1-\gamma d} - {\gamma d^2\over 6}
    {\left(\Omega_r^2 - {2\gamma\over d}\right)\over (1-\gamma d)^2},\\
  \Omega^2_\pm &\equiv& {\Omega_r^2 \mp {2\gamma\over d}
    \over 1 \mp \gamma d}, \,\,\,\,\,
  \lambda_\pm \equiv {\lambda_0\over 1\mp \gamma d}.
\end{eqnarray}
Here $\gamma_- / \gamma_+$ is of $O(\epsilon^2)$. Note that $q_-^{j}$
and the decay modes in $q_-^{(+)}$ have subradiant behavior, while
$q_+^{j}$ and the decay modes in $q_+^{(+)}$ are superradiant. For
small $d$, the time scale $\gamma_-^{-1} \gg \gamma^{-1} >
\gamma_+^{-1}\approx 1/2\gamma$, and $\gamma_-^{-1}$ goes to infinity
as $d\to 0$.

The solutions for $(\ref{eomqje5})$ and $(\ref{eomqpme5})$ with suitable
initial conditions are
\begin{eqnarray}
  q_j^{(j)}\pm \bar{q}_j^{(j)} &=& {1\over 2} e^{-\gamma_\pm t}
    \left[s_1^{\pm}e^{i\Omega_\pm t}+s_2^\pm e^{-i\Omega_\pm t}\right],\\
  q_A^{(+)}\pm q_B^{(+)} &=& {\lambda_\pm \over \Omega_\pm}
    \left( e^{-i k_1 d/2 }\pm e^{i k_1 d/2}\right)\left[
    \left(M_1^\pm-M_2^\pm\right)e^{-i\omega t}+ e^{-\gamma_\pm t}
    \left( M_2^\pm e^{i\Omega_\pm t}-M_1^\pm e^{-i\Omega_\pm t}\right)\right],
\end{eqnarray}
where $s_1^\pm \equiv [1 -\Omega_\pm^{-1}(\Omega_r+ i\gamma_\pm)]/2$,
$s_2^\pm \equiv [1 + \Omega_\pm^{-1}(\Omega_r+ i\gamma_\pm)]/2$,
$M_1^\pm \equiv (-\omega-i\gamma_\pm +\Omega_\pm)^{-1}$, and
$M_2^\pm \equiv (-\omega-i\gamma_\pm -\Omega_\pm)^{-1}$.
Actually these solutions are the zeroth-order results with $\gamma$ and
$\Omega$ replaced by $\gamma_\pm$ and $\Omega_\pm$.
So we can easily reach the simple expressions
\begin{eqnarray}
  \left<\right. Q_A^2 \left.\right>_{\rm v} &\approx&
    {\lambda_+^2\over 16\pi\gamma_+}\left[
    \left<\right. Q_A^2 \left.\right>_{\rm v}^{(0)}
    + \left<\right. Q_A, Q_B \left.\right>_{\rm v}^{(0)}
    \right]_{\gamma\to\gamma_+}^{\Omega\to\Omega_+} +
    {\lambda_-^2\over 16\pi\gamma_-}\left[
    \left<\right. Q_A^2 \left.\right>_{\rm v}^{(0)} -
    \left<\right. Q_A, Q_B \left.\right>_{\rm v}^{(0)}
    \right]_{\gamma\to\gamma_-}^{\Omega\to\Omega_-},\\
  \left<\right. Q_A,Q_B \left.\right>_{\rm v} &\approx&
    {\lambda_+^2\over 16\pi\gamma_+}\left[
    \left<\right. Q_A, Q_B \left.\right>_{\rm v}^{(0)}
    + \left<\right. Q_A^2 \left.\right>_{\rm v}^{(0)}
    \right]_{\gamma\to\gamma_+}^{\Omega\to\Omega_+} +
    {\lambda_-^2\over 16\pi\gamma_-}\left[
    \left<\right. Q_A, Q_B \left.\right>_{\rm v}^{(0)}
    -\left<\right. Q_A^2 \left.\right>_{\rm v}^{(0)}
    \right]_{\gamma\to\gamma_-}^{\Omega\to\Omega_-},
\end{eqnarray}
and so on. Here $\left<\right. .. \left.\right>_{\rm v}^{(0)}$ are
those expressions given in $(\ref{QLQRv0})$-$(\ref{PLQRv0})$ above
and in Eqs.(A9) and (A10) of Ref.\cite{LH2006} ($\left<\right. Q_A,
P_A \left.\right>_{\rm v} = \partial_t \left<\right. Q_A^2\left.
\right>_{\rm v}/2$.) The prefactors $\lambda_\pm^2/16\pi\gamma_\pm$
are put there because in our definitions for the zeroth-order results
the overall factor $\lambda_0^2$ has been expressed in terms of
$8\pi\gamma$, but now $\gamma_\pm\not= \lambda_\pm/8\pi$.

In Fig.  \ref{EntGen} we demonstrate an example in which the two
detectors are separable in the beginning but get entangled at late
times. There are three stages in their history of evolution:

1. At a very early time ($t \approx 0.15$) quantum entanglement has
been generated. This entanglement generation is dominated by the
mutual influences sourced by the initial information in the detectors
and mediated by the field. (For more early-time analysis, see
Appendix \ref{EarlyAna}.)

2. Then around the time scale $t\sim 1/\gamma_+$, the contribution from
vacuum fluctuations of the field ($\left<\right. .. \left.\right>_{\rm v}$)
takes over so that $\Sigma$ becomes quasisteady and appears to settle
down at a value depending on part of the initial data of the detectors.
More explicitly, at this stage $q_+^{(\mu)}$, $\mu =A, B, +, -$ have been
in their late-time values but $q_-^{(\mu)}$ are still about their initial
values, so
\begin{equation}
  \Sigma|_{t\sim 1/\gamma_+}\approx {\hbar^4\over 64}\left[{\sin\Omega
    d\over\Omega d}e^{-2\gamma d} + 1 -{2\over\hbar}\alpha^2\Omega \right]
  \left[ {\sin\Omega d\over \Omega d}e^{-2\gamma d} + 1 -
    {2\hbar\over\alpha^2\Omega} +{8\Lambda_1\gamma\over\pi\Omega}\right]
  \label{transSig}
\end{equation}
in the weak-coupling and short distance approximation $\gamma \ll d\Omega^2
\ll \Omega$. Here $\Sigma$ depends on $\alpha$ only. The parameter $\beta$
in initial state $(\ref{initGauss})$ is always associated with $q_+^{(j)}$
in $\left<\right. .. \left.\right>_{\rm a}$ so it becomes negligible at
this stage [cf. Eq.(25) in \cite{LCH08}]. Note that $\Sigma|_{t\sim 1/
\gamma_+}$ can be positive for small $d$ only when $\alpha$ is at the
neighborhood of $\sqrt{\hbar/\Omega}$.

3. The remaining initial data persist until a much longer time scale
$t\sim 1/\gamma_-$ when $\Sigma$ approaches a value consistent with
the late-time results given in Sec. \ref{resient}, which are
contributed purely by the vacuum fluctuations of the field and
independent of any initial data in the detectors. In this example the
detectors have residual entanglement, though small compared to those
in stage 2.

The above behaviors in stages 2 and 3 cannot be obtained by 
including only the first-order correction from the mutual influences.
Thus in this example we conclude that the mutual influences of the
detectors at very short distance generate a transient entanglement
between them in midsession, 
while vacuum fluctuations of the field with the mutual influences included
give the residual entanglement of the detectors at late times.

For the detectors initially entangled, only the early-time behavior looks
different from the above descriptions. Their entanglement dynamics are
similar to the above in the second and the third stages.

\section{Discussion}

\subsection{Physics represented by length scales}

The physical behavior of the system we studied may be characterized
by the following length scales:

{\it Merge distance $d_{min}$ in Eq.}(\ref{dmin}).--Two detectors
separated at a distance less than $d_{min}$ would be viewed as those
located at the same spatial point;

{\it Radius of instability $d_{ins}$ in Eq.}(\ref{dins}).--For 
any two detectors at a distance less than $d_{ins}$, their mode
functions will grow exponentially in time so the quantum fluctuations of
the detector diverge at late times;

{\it Entanglement distance $d_{ent}$ in Eq.}$(\ref{dentdef})$.--Two 
detectors at a distance less than $d_{ent}$ will be entangled at
late times, otherwise separable;

And $d_0$ defined in Sec. \ref{0bad}.--For $d < d_0$ the zeroth-order
results breakdown.
A stable theory should have $d_{ent}$ and $d_{min}$ greater than $d_{ins}$.

\subsection{Direct interaction and effective interaction}

In a closed bipartite system a direct interaction between the two
parties, no matter how weak it is, will generate
entanglement at late times. However, as we showed above, an effective
interaction between the two detectors mediated by quantum fields will
not generate residual entanglement (though creating transient
entanglement is possible) if the two detectors are separated far
enough, where the strength of the effective interactions is weak but
not vanishing.

\subsection{Comparison with 2HO QBM results}

When $d \to d_{min}$ with large enough $\Omega$, our model will
reduce to a 2HO QBM model with real renormalized natural frequencies
for the two harmonic oscillators. Paz and Roncaglia \cite{PR07} have
studied the entanglement dynamics of this 2HO QBM model and found
that, at zero temperature, for both oscillators with the same natural
frequency, there exists residual entanglement at late times in some
cases and infinite sequences of sudden death and revival in other
cases. In the latter case the averaged asymptotic value of negativity
is still positive and so the detectors are ``entangled on average."

While our results show that the late-time behavior of the detectors
is independent of the initial state of the detectors, the asymptotic
value of the negativity at late times in \cite{PR07} does depend on
the initial data in the detectors (their initial squeezing factor).
This is because in \cite{PR07} the two oscillators are located
exactly at the same point, namely, $d=0$, so $\gamma_- =0$ and the
initial data carried by $q_-^{(j)}$ persists forever. Since in our
cases $d$ is not zero, the ``late" time in \cite{PR07} actually
corresponds to the time interval with $(1/\gamma_+) \ll t \ll
(1/\gamma_-)$ in our cases, which is not quite late for our
detectors.

\subsection{Where is the spatial dependence of entanglement coming from?}

Two factors are responsible for the spatial dependence of
entanglement. The first one is the phase difference of vacuum
fluctuations that the two detectors experience. This is mainly
responsible for the entanglement outside the light cone in all
coupling strengths and those inside the light cone
with sufficiently large separation in the weak-coupling limit,
such as the cases in Sec.\ref{ZOR}.
The second factor is the interference of retarded mutual
influences, which are generated by backreaction from the detectors
to the field. It is important in the cases with small separation
between the detectors, such as those in Sec.\ref{createEnt}.

\subsection{Non-Markovian behavior and strong coupling}

In our prior work \cite{LH2006} and \cite{LCH08}, the non-Markovian
behavior arises mainly from the vacuum fluctuations experienced by
the detectors, and the essential temporal nonlocality in the
autocorrelation of the field at zero temperature manifests fully in
the strong-coupling regime. Nevertheless, in Sec. \ref{createEnt} one
can see that, even in the weak-coupling limit, once the spatial
separation is small enough and the evolution time is long enough, the
mutual influences will create some non-Markovian behavior very
different from those results obtained from perturbation theory with
higher-order mutual influences on the mode functions neglected.\\

\noindent{\bf Acknowledgement} S.Y.L. wishes to thank Jen-Tsung Hsiang
for helpful discussions. This work is supported in part by grants
from the NSF Grants No. PHY-0426696, No. PHY-0601550, No. PHY-0801368,
and the Laboratory for Physical Sciences.

\begin{appendix}

\section{Late-time analysis on mode functions}
\label{LateAna}

\begin{figure}
\includegraphics[width=6cm]{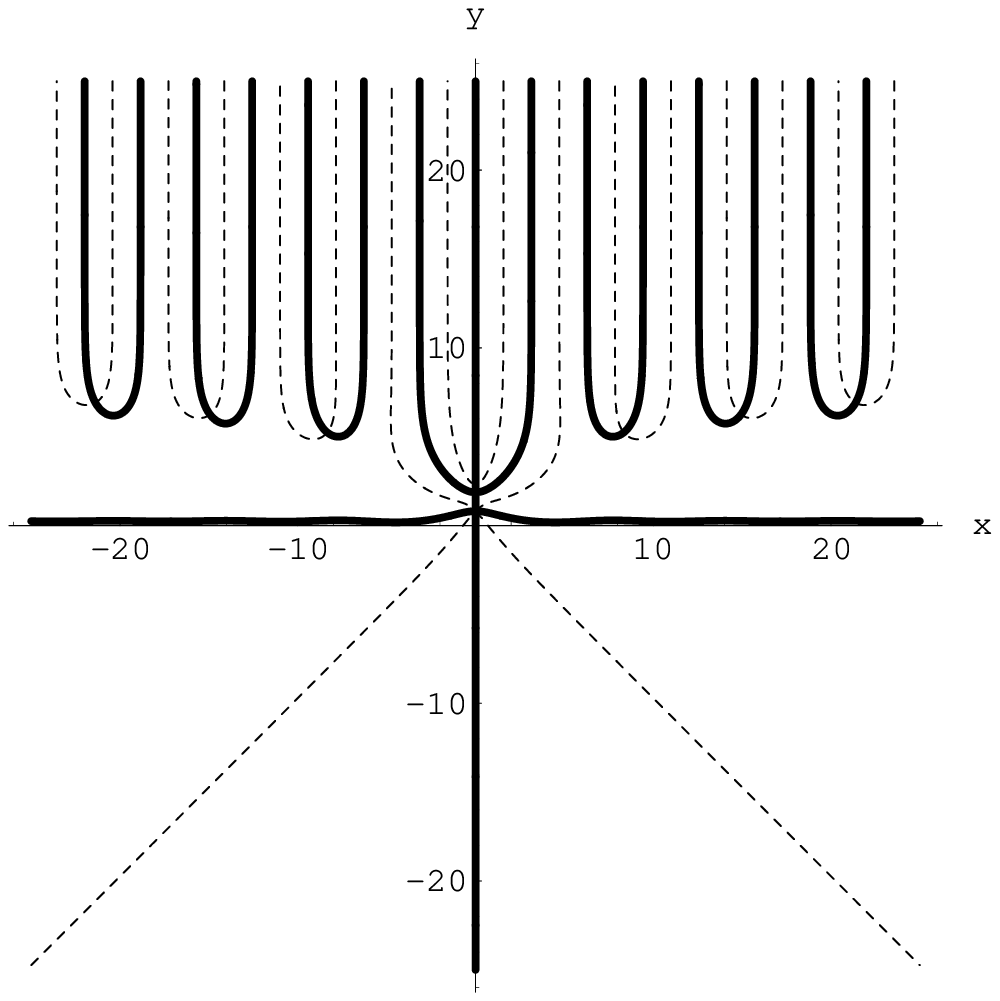}
\includegraphics[width=6cm]{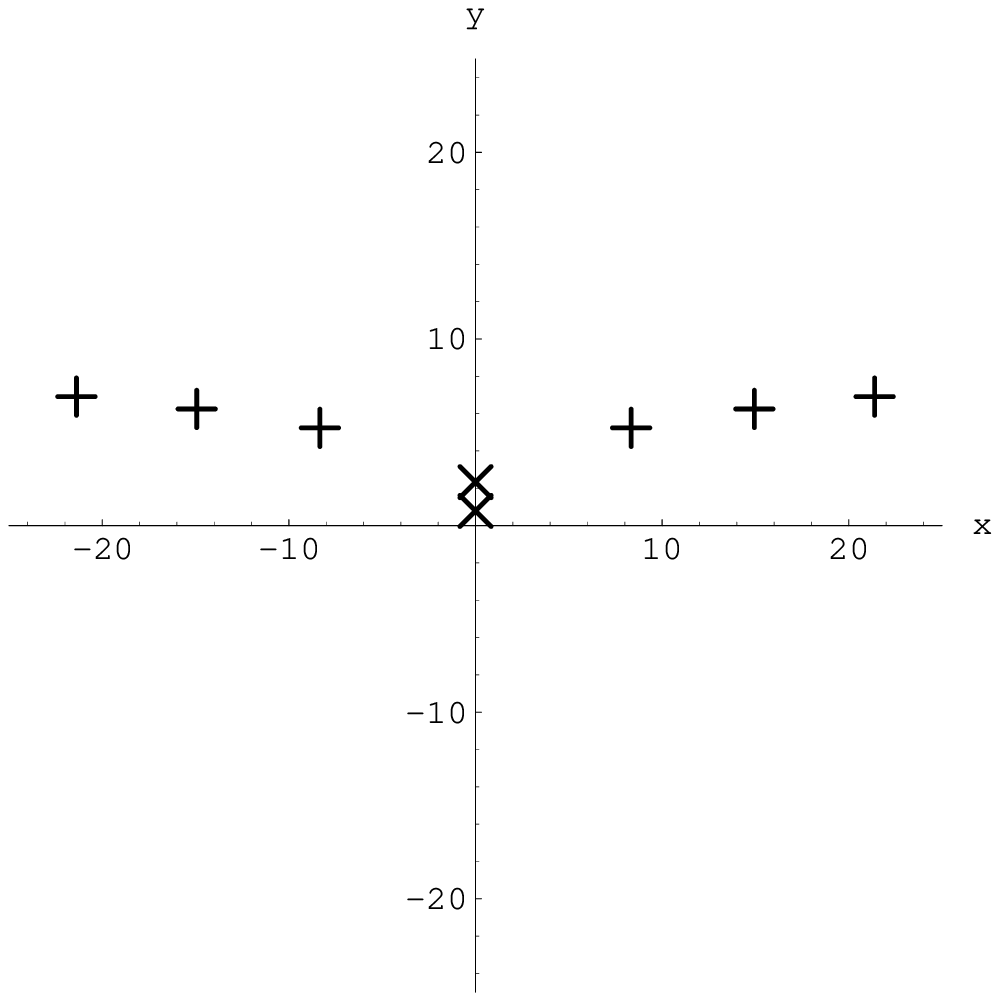}\\
\includegraphics[width=6cm]{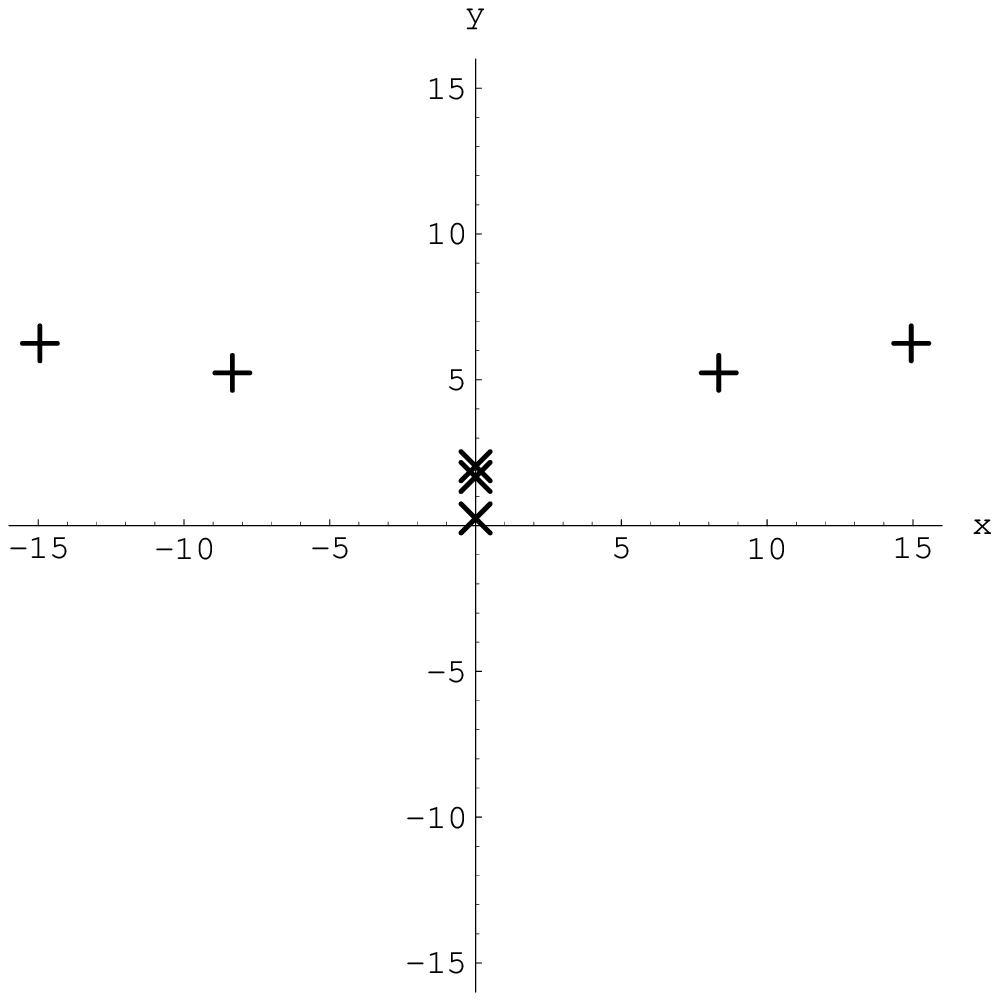}
\includegraphics[width=6cm]{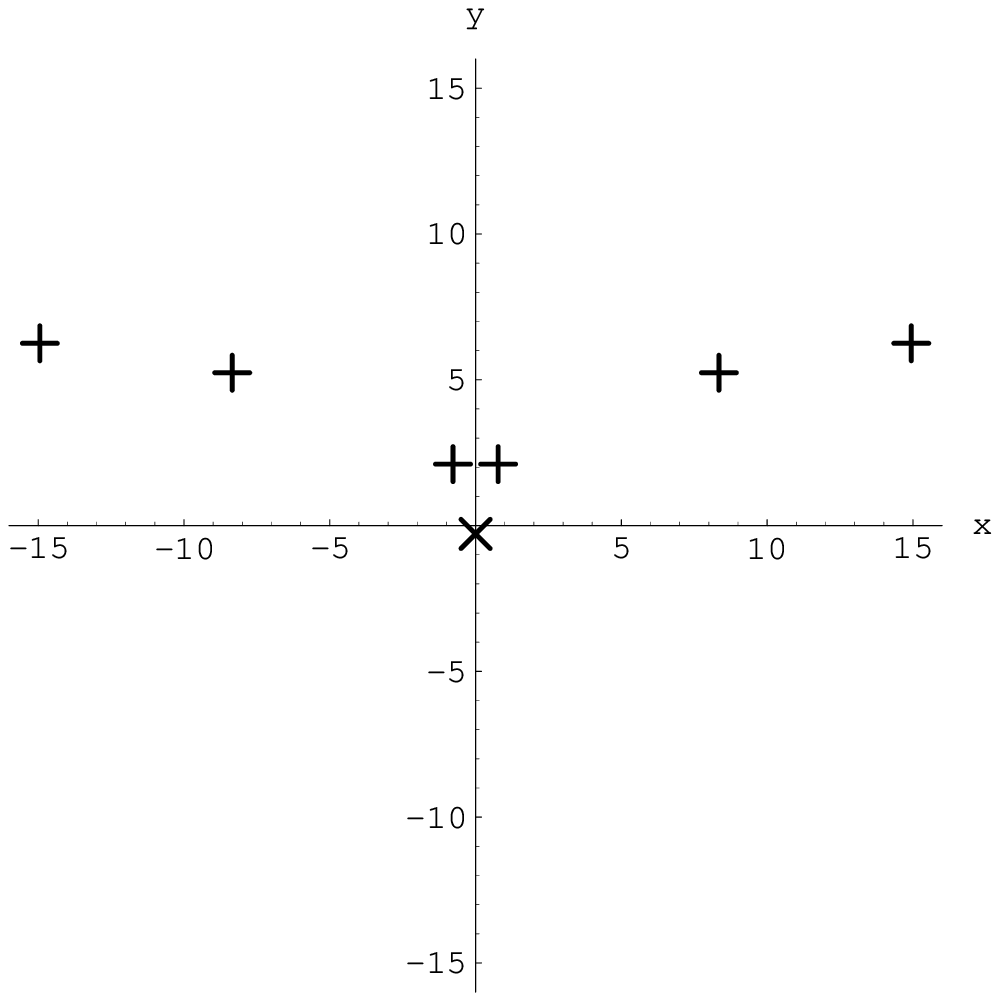}
\caption{
(Upper left) The solutions to ($\ref{eomK}$) for the complex frequency
$K_j=x_j+iy_j$ of $q_+^{(A)}$ [defined in $(\ref{qR+FT})$] are located
at the intersections of the dashed and solid curves, which represent
the solutions to Eqs. $(\ref{eomRe})$ and $(\ref{eomIm})$, respectively.
Here $\gamma=0.25$, $\Omega=0.9000202$, $d=1$. (Upper right) The same
case, but here ``$+$" denotes complex solutions and ``$\times$" denotes
purely imaginary solutions for $K$. There are two purely
imaginary solutions for $K$ in this case. (Lower left) There are three
purely imaginary solutions for $K$ when $\gamma=0.25$, $\Omega=0.8$, $d=1$.
(Lower right) Solutions for $K$ when $\gamma=0.25$, $\Omega=0.3$, $d=1$.
There is only one purely imaginary solution, which is located in the 
lower half of the complex $K$ plane.}
\label{SolK}
\end{figure}

Let
\begin{equation}
  q_+^{(A)} (t) = \sum_j c_j e^{i K_j t}, \label{qR+FT}
\end{equation}
Equation (\ref{eomqpm}) gives
\begin{equation}
  \sum_j c_j\left[-K_j{}^2+ 2i\gamma K_j + \Omega_r^2 \right]
  e^{i K_j t}= {2\gamma\over d} \sum_{j'}c_{j'} e^{i K_{j'}(t-d)} .
\end{equation}
At late times, one is allowed to perform the 
Fourier transformation on both sides with $t$ integrations over
$(-\infty, \infty)$ to obtain
\begin{equation}
  -K_j^2+ 2i\gamma K_j + \Omega_r^2 = {2\gamma\over d} e^{-i K_j d} .
  \label{eomK}
\end{equation}
There are infinitely many solutions for $K_j$ in the complex $K$
plane, so one needs infinitely many initial conditions to fix the
factors $c_j$. Our $q_+$ chosen as a free oscillator at the initial
moment and unaffected by its own history until $t=d$ in principle can
be specified by a set of $c_j$'s. Suppose this is true. Writing $K_j
\equiv x_j + i y_j$, the real and imaginary parts of $(\ref{eomK})$
then read
\begin{eqnarray}
  (y-\gamma)^2 -x^2 + \Omega^2 &=& {2\gamma\over d} e^{y d}\cos x d,
    \label{eomRe} \\
  x (y -  \gamma) &=& {\gamma\over d} e^{y d}\sin x d. \label{eomIm}
\end{eqnarray}
The solutions for them are shown in Fig.  \ref{SolK}. The left-hand
side of $(\ref{eomRe})$ is a saddle surface over the $xy$ space,
while the right-hand side of $(\ref{eomRe})$ is exponentially growing
in the $+y$ direction and oscillating in the $x$ direction. For
$(\ref{eomIm})$, the situation is similar. From Fig.  \ref{SolK}, one
can see that there is no complex solution for $K$ with nonvanishing
real part and negative imaginary part ($x\not=0$ and $y\le 0$). The
solutions for $K$ with its imaginary part negative must be purely
imaginary. Indeed, from $(\ref{eomIm})$ and Fig.  \ref{SolK}
(upper right), one sees that
when $x\not= 0$, if $y\le 0$, then $(y-\gamma)\le -\gamma$, but $-0.2172
\gamma  \alt \gamma e^{y d}(\sin x d)/(x d) < \gamma $, so there is no
solution of $(\ref{eomIm})$ with $y\le 0$ and $x\not=0$.

When $\Omega_r^2 > 2\gamma/d$, one finds that all solutions for $K$ in
$(\ref{eomK})$ are located in the upper half of the complex $K$ plane,
{\it i.e.}, all $y_j>0$, which means that all modes in $(\ref{qR+FT})$
decay at late times.

When $\Omega_r^2 = 2\gamma/d$, there exists a solution $K=0$, with
other solutions on the upper half $K$ plane. This implies that
$q_{+}^{(A)}$ becomes a constant at late times.

When $\Omega_r^2 < 2\gamma/d$, there must exist one and only one solution
for $K$ with negative $y$, which corresponds to the unstable growing mode.
This is consistent with our observation in Sec.\ref{instab}.

Therefore, we conclude that $q_+^{(A)}$ is stable and decays at late
times only for $\Omega_r^2 > 2\gamma/d$.

As for $q_-^{(A)}$,
from $(\ref{EOMqR-})$ it seems that $q_-^{(A)}$ would oscillate at
late times. However, similar analysis gives the conclusion that
$q_-^{(A)}$ decays at late times for all cases. Thus, by symmetry,
all $q_{j}^{(i)}$ decay at late times in the stable regime
$\Omega_r^2 > 2\gamma/d$.

Now we turn to $q^{(+)}_{A,B}$.  Equation $(\ref{eomqA+2})$ implies that
\begin{eqnarray}
  \left( \partial_t^2 +2\gamma \partial_t + \Omega_r^2 \right)^2
    q_B^{(+)}(t,{\bf k}) &=&
  \left({2\gamma\over d}\right)^2 q_B^{(+)}(t-2d,{\bf k})+ \nonumber\\
  & & {\lambda_0}e^{-i\omega t}\left[\left( -\omega^2-2\gamma\omega
    +\Omega_r^2\right)e^{i k_1 d/2} + {2\gamma\over d}
    e^{i\omega d -i k_1 d/2}\right],
\end{eqnarray}
at late times. Again, let
\begin{equation}
  q^{(+)}_{B} (t,{\bf k}) = \sum_j c^j_{\bf k} e^{i K^j_{\bf k} t},
\end{equation}
then one has
\begin{eqnarray}
  \sum_j c^j_{\bf k} \left[ -\left(K_{\bf k}^j\right)^2 +
    2i\gamma K_{\bf k}^j + \Omega_r^2 \right]^2 e^{i K^j_{\bf k} t} &=&
   \sum_j c^j_{\bf k}\left({2\gamma\over d}\right)^2 e^{i K^j_{\bf k}(t-2d)}+
   \nonumber\\ & & {\lambda_0}e^{-i\omega t}\left[\left( -\omega^2-
     2i\gamma\omega +\Omega_r^2\right)e^{i k_1 d/2} + {2\gamma\over d}
    e^{i\omega d -i k_1 d/2}\right].
\end{eqnarray}
After a Fourier transformation, for $K^j_{\bf k}\not= -\omega$,
the above equation becomes
\begin{eqnarray}
   \left[ -\left(K_{\bf k}^j\right)^2 +
    2i\gamma K_{\bf k}^j + \Omega_r^2 \right]^2  &=&
   \left({2\gamma\over d}\right)^2 e^{-2 i K^j_{\bf k}d},
\end{eqnarray}
which is the square of Eq.$(\ref{eomK})$ for $q_+^{(A)}$, or the square 
of the counterpart for $q_-^{(A)}$. So these $K^j_{\bf k}$ modes decay at 
late times for $\Omega_r^2 > 2\gamma/d$ as $q_+^{(A)}$ and $q_-^{(A)}$ do.
On the other hand, if, say, $K^0_{\bf k} = -\omega$, one has
\begin{eqnarray}
    \left[ -\omega^2 + 2i\gamma \omega + \Omega_r^2 \right]^2 c^0_{\bf k}
     &=& \left({2\gamma\over d}\right)^2 c^0_{\bf k}e^{-2i\omega d}+
   \nonumber\\ & & {\lambda_0}\left[\left( -\omega^2-2i\gamma\omega
    +\Omega_r^2\right)e^{i k_1 d/2} + {2\gamma\over d}
    e^{i\omega d -i k_1 d/2}\right].
\end{eqnarray}
This equation will not hold unless
\begin{equation}
  c^0_{\bf k} = {{\lambda_0}\left[\left( -\omega^2-2i\gamma\omega
    +\Omega_r^2\right)e^{i k_1 d/2} + {2\gamma\over d}
    e^{i\omega d -i k_1 d/2}\right]\over
    \left[ -\omega^2 + 2i\gamma \omega + \Omega_r^2 \right]^2 -
    \left({2\gamma\over d}\right)^2 e^{-2i\omega d}}. \label{lateTc0}
\end{equation}
Therefore, for $\Omega_r^2 > 2\gamma/d$, the only mode which survives
at late times will be $e^{-i\omega t}$, and
\begin{equation}
  q^{(+)}_B (t,{\bf k})|_{t\gg 1/\gamma} = c^0_{\bf k}e^{-i\omega t}.
  \label{lateTqp}
\end{equation}
This is nothing but the sum of the $e^{-i\omega (t-nd)}$ part in Eq.
$(\ref{qjp})$ with $t \to \infty$ so summing from $n=0$ to $\infty$.
Thus, $(\ref{lateTqp})$ with $(\ref{lateTc0})$ has included the
mutual influences to all orders. The above analysis also indicates
that the $e^{-\gamma (t-nd)}$ part in $(\ref{qjp})$ really decays at
late times for $\Omega_r^2 > 2\gamma/d$.

\section{Early-time behaviors in weak-coupling limit}
\label{EarlyAna}

\begin{figure}
\includegraphics[width=7cm]{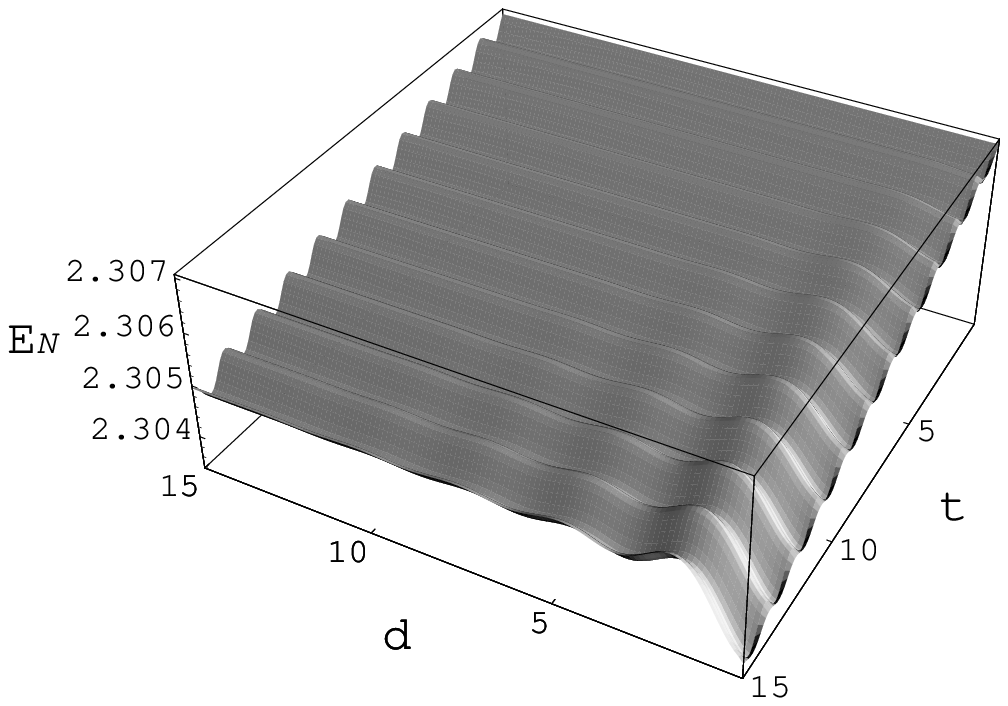}
\includegraphics[width=5cm]{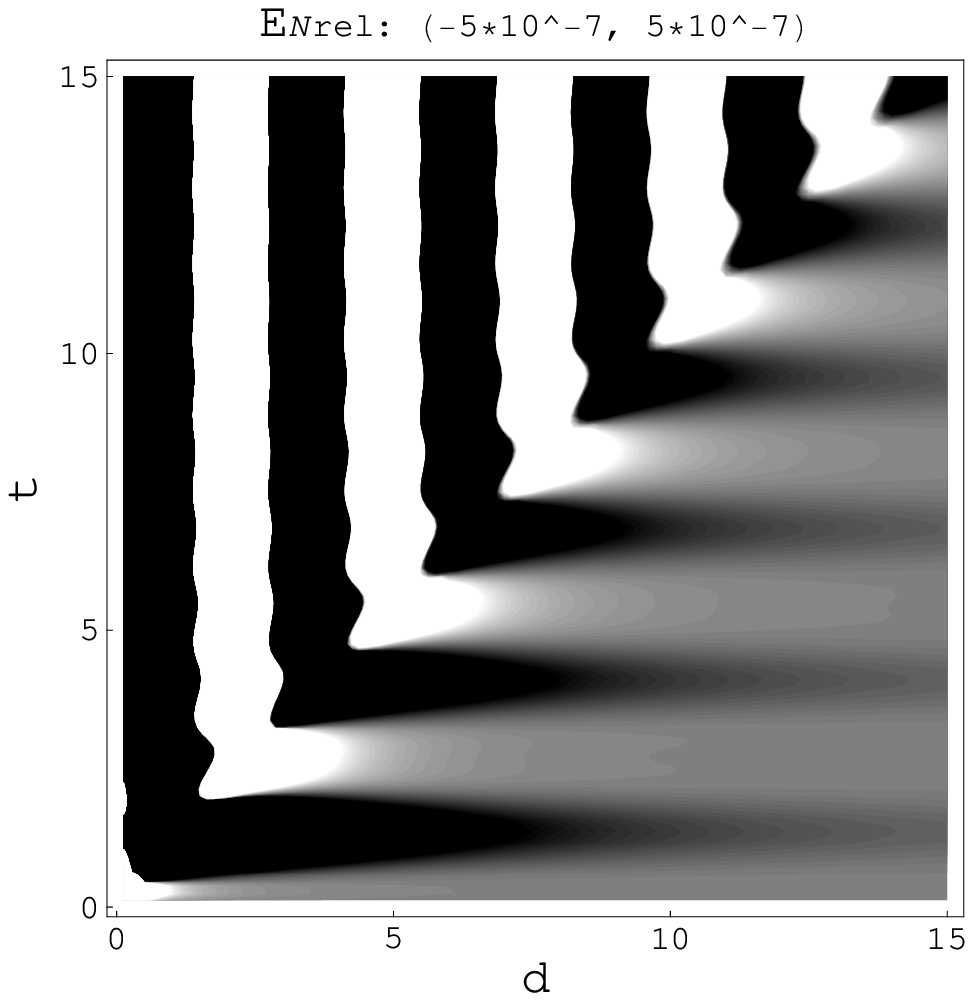}\\ 
\includegraphics[width=7cm]{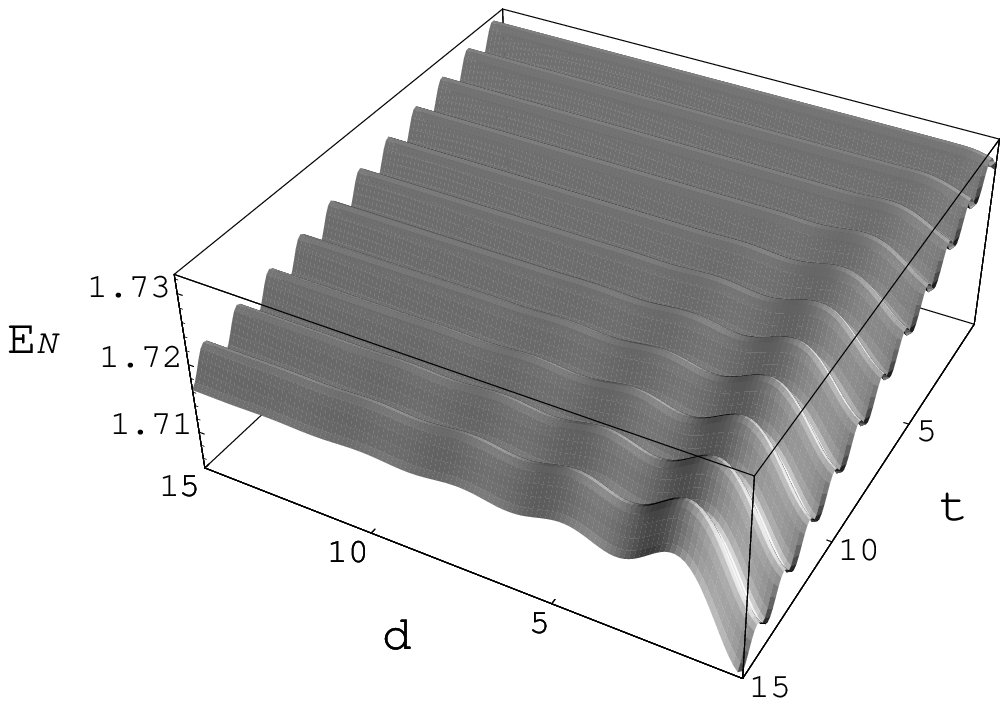}
\includegraphics[width=5cm]{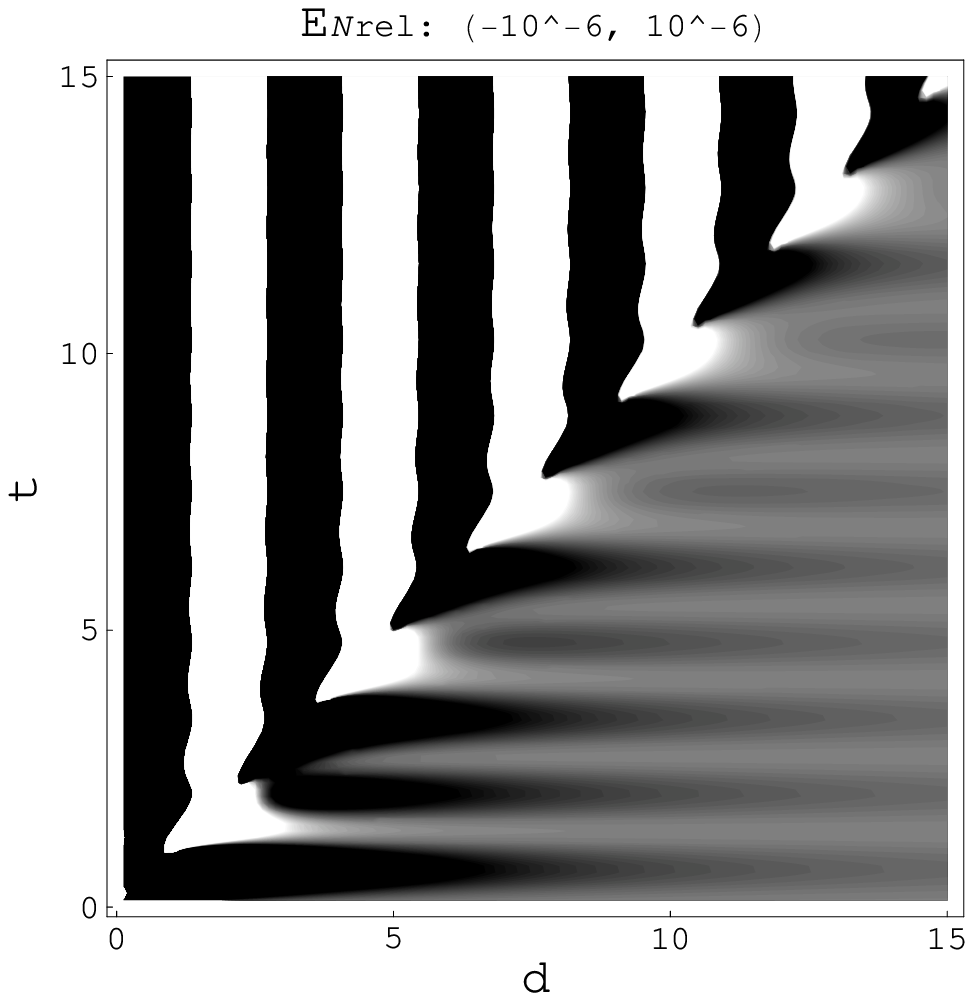}
\caption{The early-time evolution of $E_{\cal N}$ with the first-order
mutual influence included for different initial states of the detectors
with $1/8 < d < 15$.
(Upper row) All parameters are the same as those in Fig. \ref{zeroS}
where $(\alpha,\beta)=(1.1,4.5)$ and the initial state of the detectors
is entangled. Compared with Fig. \ref{zeroS}, one can see that
the distortion of the interference pattern due to the mutual influences
is tiny.
(Lower row) $(\alpha,\beta)=(1.5, 0.2)$, the detectors also initially
entangled. The distortion by the mutual influences is also tiny.
As indicated by Eq. $(\ref{ENrel})$, 
the complicated structure of $E_{{\cal N}{\rm rel}}$ outside the
light cone is reducing to simple oscillations as time goes larger.}
\label{firstS2}
\end{figure}

\begin{figure}
\includegraphics[width=5cm]{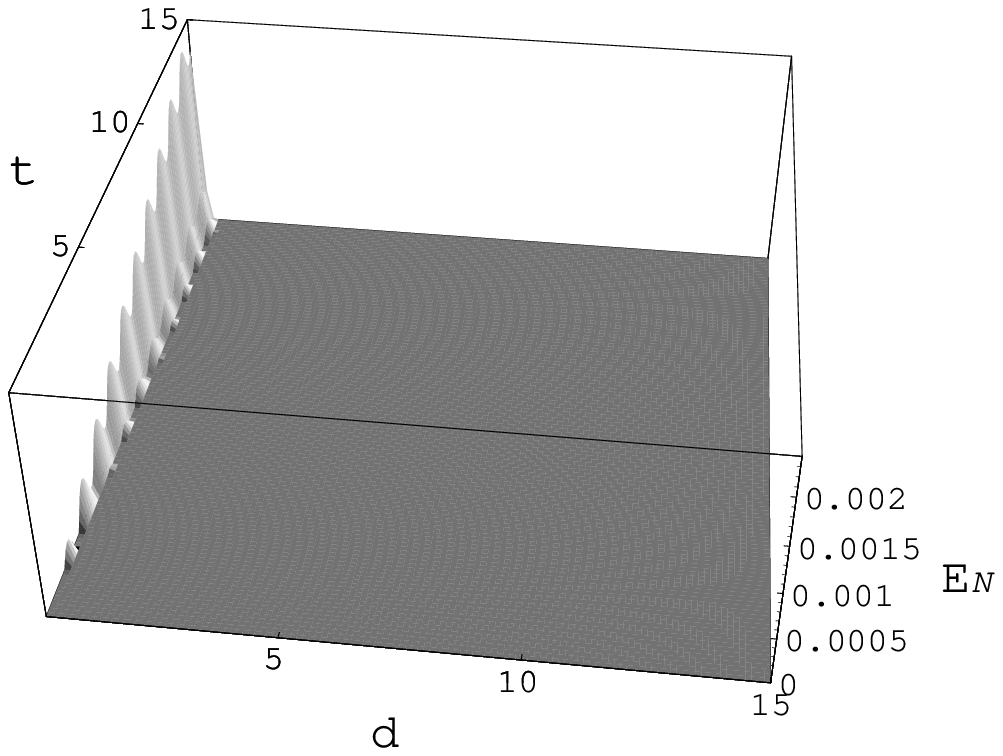}
\includegraphics[width=6cm]{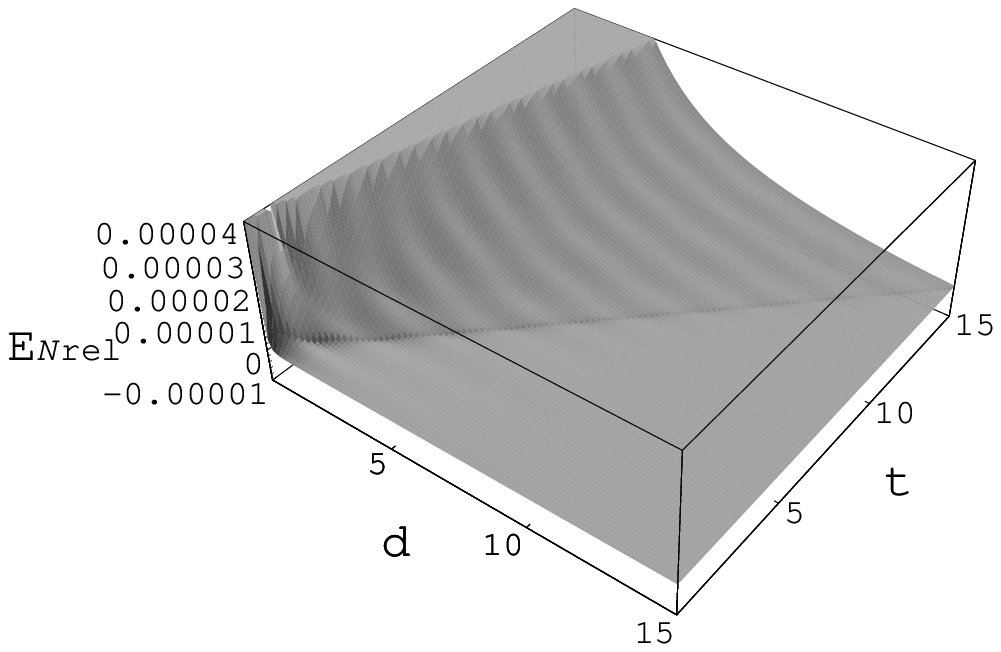}
\caption{The early-time evolution of $E_{\cal N}$ for an initially
separable detector pair with the first-order mutual influence included.
The parameters are the same as those in Fig. \ref{zeroS}
except $(\alpha,\beta)=(1,1)$ and $1/15 < d < 15$ here.
From the left plot, one finds that quantum entanglement is created at
small $d$ due to the mutual influences. In the right plot, one can
see that there is no clear interference pattern in $d$ similar to those
in Fig. \ref{zeroS} for $E_{{\cal N}{\rm rel}}$ inside the light cone.
Note that here $E_{{\cal N}{\rm rel}}\equiv -\log_2 2c_-(t,d) + \log_2
2c_-(t,\infty)$ instead of $E_{\cal N}(t,d)- E_{\cal N}(t,\infty)$.
The detectors with smaller separation $d$ always get a greater value
of $-\log_2 2c_-$. 
While there are small oscillations
outside the light cone, the smaller separation $d$ always associates the
greater value of $E_{{\cal N}{\rm rel}}$. The amplitude of the small
oscillation is the same order of $\gamma^2\Lambda_0$ and 
$\gamma^2\Lambda_1$.}\label{firstS1}
\end{figure}

In the weak-coupling limit,
the cross correlators $\left<\right.{\cal R}_A, {\cal R}'_B \left.
\right>$ with ${\cal R}, {\cal R}'= Q,P$ are small until one
detector enters the other's light cone. From this observation one might
conclude that the cross correlations between the two detectors are
mainly generated by the mutual influences sourced by the quantum state
of the detectors and mediated by the field. This is not always true.

As shown in Sec.\ref{Earlyd<t}, the interference pattern inside the
light cone has been there in the zeroth-order results, where the mutual
interferences on the mode functions are not included. A comparison of
the first-order results in the upper plots in Fig. \ref{firstS2} and
those of the zeroth-order in Fig. \ref{zeroS} shows that the corrections
to entanglement dynamics from mutual influences at early times are
pretty small in that case. Actually the early-time dynamics of entanglement
in both examples in Fig. \ref{firstS2} are dominated by the zeroth-order
results, thus by the phase difference of vacuum fluctuations in
$\left<\right.{\cal R}_A,{\cal R}'_B\left. \right>_{\rm v}^{(0)}$ rather
than mutual influences. One can see this explicitly by inserting the
mode functions in the weak-coupling limit with the first-order correction
from the mutual influences into Eq.(25) in Ref. \cite{LCH08}, and write
\begin{eqnarray}
  \Sigma(t) \approx \Sigma_0 + \sigma_1^{(0)} t + \sigma_2^{(0)} t^2 +
    \theta(t-d)\left[\sigma_1^{(1)} (t-d)+ \sigma_2^{(1)} (t-d)^2\right]  +
    O(\gamma^3) \label{earlySig}
\end{eqnarray}
at early times when $O(e^{-\gamma_e -(\Lambda_0/2)}/\Omega) < t \ll
O(1/\gamma\Lambda_i)$, $i=0,1$. Here $\Sigma_0$, $\sigma_1$, and $\sigma_2$
depend on $\alpha$, $\beta$ and of $O(\gamma^0)$, $O(\gamma)$, and $O(\gamma^2)$,
respectively. Then it is easy to verify that mutual influences are negligible in
the dominating $\sigma_1^{(1)}$ term after $\theta(t-d)$ for the initial
states with the value of $\beta^2$ not in the vicinity of $\hbar^2/\alpha^2$
or $\alpha^2\Omega^2$.

In contrast, if the initial state ($\ref{initGauss}$) is nearly separable
($\beta^2 \approx \hbar^2/\alpha^2$), mutual influences will be important
in the detectors' early-time behavior.
In this case, dropping all terms with small oscillations in time, the
factors in $(\ref{earlySig})$ are approximately
\begin{eqnarray}
  \Sigma_0 &\approx& {\hbar^2\over 4\pi^2\alpha^4\Omega^4}
    \left[\hbar^2\gamma\Lambda_1 + \alpha^4 \Omega^2 \gamma
    (2\Lambda_0+\Lambda_1)\right]^2, \nonumber\\
  \sigma_2^{(0)} &\approx& {\gamma^2\hbar^2\left(\hbar -
    \alpha^2\Omega\right)^4\over 4\Omega^2\alpha^4},\nonumber\\
  \sigma_2^{(1)} &\approx& -{\gamma^2\hbar^2\left(\hbar^2 -
    \alpha^4\Omega^2\right)^2\over 4\Omega^4\alpha^4 d^2}, \nonumber\\
  \sigma_1^{(0)} &\approx& {\gamma\hbar^2\over 2\pi \Omega^3}\left[
    2\Omega^2\gamma\Lambda_0 +\left({\hbar^2\over\alpha^4}+\Omega^2\right)
    \gamma\Lambda_1\right]\left(\hbar -\alpha^2\Omega\right)^2,
\end{eqnarray}
with $\sigma_1^{(1)}$ negligible. So $\Sigma$ evolves as the following.
In a very short-time scale $O(e^{-\gamma_e-(\Lambda_0/2)}/\Omega)$
after the interaction is switched on, $\Sigma$ jumps from its initial
value $(\approx 0)$ to a value of the same order of $\Sigma_0$, which is
positive and determined by the numbers $\Lambda_0$ and $\Lambda_1$
corresponding to the cutoffs of this model (the difference
from the exact value is due to the oscillating terms dropped).
For $\alpha^2\not=\hbar/\Omega$ so $Q_A$ and $Q_B$ are each in a
squeezed state initially, the detectors keep separable at $t\le d$
since $\sigma_1^{(0)}$ and $\sigma_2^{(0)}$ are positive definite. But
$\sigma_2^{(1)}$ is negative and proportional to $1/d^2$, thus after
entering the light cone of the other detector, if the separation $d$ is
sufficiently small, or
\begin{equation}
  d < d_1 \equiv {1\over\Omega} \left| \hbar +\alpha^2\Omega\over \hbar
    -\alpha^2\Omega\right|, \label{defd1}
\end{equation}
$\sigma_2^{(1)}$ can overwhelm $\sigma_2^{(0)}$ and alter the evolution
of $\Sigma$ from concave up to concave down in time. If this happens,
the quantity $\Sigma$ could become negative after a finite
``entanglement time"
\begin{equation}
  t_{ent} \approx {1\over 2}\left|\sigma_2^{(0)} + \sigma_2^{(1)}\right|^{-1}
    \left[\sigma_1^{(0)} - 2 \sigma_2^{(1)} d + \sqrt{\left(\sigma_1^{(0)} -
    2 \sigma_2^{(1)} d\right)^2+ 4\left|\sigma_2^{(0)} + \sigma_2^{(1)}\right|
    \left(\Sigma_0+ \sigma_2^{(1)} d^2\right)}\right].
\end{equation}
This explains the entanglement generation at small $d$ in Fig. \ref{firstS1}.
[Note that the above prediction could fail if $t_{ent}>O(1/\gamma\Lambda_i)$,
$i=0,1$, and even for $t_{ent} < O(1/\gamma\Lambda_i)$ the above estimate
on $t_{ent}$ could have an error as large as $O(2 \pi/\Omega)$ due to the
dropped oscillating terms.]
The first-order corrections to $\left<\right. .. \left.\right>_{\rm a}$
contribute the $\sigma_2^{(1)}\cos^2\Omega d$ part of $\sigma_2^{(1)} =
\sigma_2^{(1)}(\cos^2\Omega d+\sin^2\Omega d)$, so for those cases with
separations small enough such that $\sin^2 \Omega d \ll \cos^2\Omega d$ the
early-time entanglement creations are mainly due to mutual influences of
the detectors, which is causal.

$d_1$ in $(\ref{defd1})$ can serve as an estimate for the maximum distance
that transient entanglement can be generated from a initially separable
state in the weak-coupling limit,
while for the detectors with the spatial separation between $d_1$ and
$d_{ent}$ the transient entanglement generated at early times will
disappear at late times.

\end{appendix}

\end{document}